\documentclass[twocolumn]{aa}
\usepackage{graphicx}
\usepackage[varg]{txfonts}
\usepackage{subcaption}
\newcommand{\lta}{\mathrel{\hbox{\raise 0.6 ex \hbox{$<$}\kern
                   -1.8 ex\lower .5 ex\hbox{$\sim$}}}}
\newcommand{\gta}{\mathrel{\hbox{\raise 0.6 ex \hbox{$>$}\kern
                   -1.7 ex\lower .5 ex\hbox{$\sim$}}}}
\DeclareMathAlphabet{\mathsc}{T1}{cmr}{m}{sc}

\begin{document}

\title{Zeeman Doppler mapping \thanks{Instead of using the terms
    Zeeman Doppler imaging (ZDI) or magnetic Doppler imaging (MDI), we
    respect the original meaning given by \citet{SemelSe1989} to ZDI,
    knowing that he always objected to its use for Doppler mapping.} deconstructed}

\author{M.J.~Stift\inst{\ref{inst1},} \and F.~Leone\inst{\ref{inst2}}}

\institute{
     Kuffner-Sternwarte, Johann~Staud-Strasse 10,  A-1160~Wien, Austria\label{inst1}
\and Dipartimento di Fisica e Astronomia, Universit{\`a} di Catania,
     Sezione Astrofisica, Via S. Sofia 78, I-95123 Catania, Italy\label{inst2}
  }

\date{submitted for publication in A\&A 2025}

\abstract{
  {\em Aims}. Magnetic and abundance maps of chemically peculiar (CP) stars,
    derived with the help of Zeeman Doppler mapping, have invariably been used
    as arguments against theories, in particular atomic diffusion theory. We
    intend to expose the fallacy of these claims.
  \\
  {\em Methods}. We have identified in the literature those (5) CP stars
    for which multiple maps have been published, all based on the same Zeeman
    Doppler mapping strategy. For each of these stars we have then carried
    out inter comparisons between the recovered distributions of magnetic
    field and of abundances.
  \\
  {\em Results}. Agreement between maps often turns out to be quite poor in
    regard to both abundances, field topology and absolute field strengths.
    Maps based on the same set of observations can differ considerably, even
    when they are coming from the same authors.
  \\
  {\em Conclusions}. It becomes clear that Zeeman Doppler mapping cannot be
    guaranteed to yield unique results. When a number of physically impossible
    magnetic geometries all provide good fits to the observed Stokes $IQUV$
    profiles, these solutions must necessarily be spurious and cannot be used
    as constraints to diffusion theory.
  }

\keywords{
  Stars: chemically peculiar -- Stars: magnetic field -- Stars: abundances
  --  Methods: numerical
  }

\maketitle

\section{Introduction}
\label{intro}

In the 1940s, a former student at Uppsala University, Hannes
Alfv{\'e}n, developed the theory of magneto-hydrodynamics (MHD),
for which he received the Nobel Price in Physics in 1970. This
theory provided insight, among others, into the behaviour
of strongly magnetic structures in the solar photosphere
(sunspots) and the loops and arcades in the highly ionised
corona. It seemed however to bear little relevance to magnetic
upper main sequence chemically peculiar (CP) stars, since the
magnetic geometries of these stars were thought of and modelled
in terms of low order potential fields. When Doppler mapping, as
introduced largely in its present form by \citet{VogtVoPeHa1987},
was extended to Zeeman Doppler mapping (ZDM) of CP stars by
\citet{KochukhovKoPi2002} the question of force-free fields was
completely overlooked, never to be discussed for a full 2 decades.
Unnoticed by the ZDM community, \citet{BraithSpruit2017}
mentioned the fact that already at field strengths of 300\,G,
as in the weakest-field CP stars, magnetic pressure equals gas
pressure at the photosphere, i.e. $\beta = 8\,\pi\,P / B^2 = 1$.
Substantially stronger fields in CP stars therefore must be
force-free to be stable over years and decades. Some time later,
\citet{StiftLeone2022} realised that the formulae underlying
modern ZDM inversions were incompatible with the (very) strong
fields observed in most target CP stars, as they violated the
force-free condition.

The analysis of $\alpha^2$\,CVn by \citet{KochukhovKoPiIlTu2002},
based on Stokes $IV$ only, employed a second-order multipolar
expansion. Two years later, in the study of 53\,Cam,
\citet{KochukhovKoBaWaetal2004} abandoned this approach,
assuming that unique recovery of a stellar magnetic map is
possible with high quality observational data in $IQUV$,
obviating any a priori assumptions about the magnetic geometry.
The results were sobering, as the resulting field did not even
prove divergence-free.

The decisive step away from force-free fields was made by
\citet{KochukhovKoWa2010}. Given the long-term (decades) stability
of the magnetic fields of CP stars, potential fields
\citep{Jardineetal1999} should have been the choice on account
of the absence of currents (see e.g. \citealt{WinchWiIvTuSt2005}).
They however sidestepped potential fields, choosing the set of
formulae given by \citet{Donatietal2006} instead, which include
toroidal components and assume decoupling between radial and
horizontal poloidal field components. As shown by
\citet{StiftLeone2022}, this added freedom (3 times the number
of free parameters) violates the force-free condition.

From these considerations one far-reaching conclusion can be
drawn: at variance with all previous claims concerning the
alleged uniqueness of ZDM inversions, one is faced with the
fact that there exist multiple solutions. Indeed, when magnetic
maps yield good fits to all 4 Stokes parameters, although they
are physically absolutely impossible, one cannot uphold the
fiction of uniqueness. This logically leads to the next question,
viz. whether ZDM offers even more spurious solutions for a given
star. We will investigate this in detail in what follows.

\begin{figure}
\includegraphics[height=90mm, angle=270]{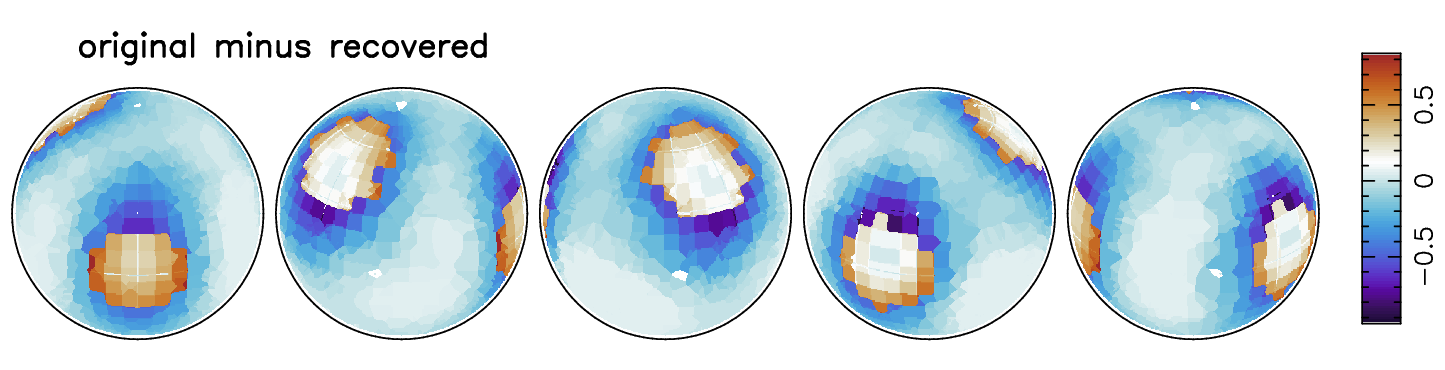}
\caption{
  Abundance differences between the original 2002 3-spot map and
  the recovered map.
}
\label{orig_maps}
\end{figure}

\section{The (Zeeman) Doppler mapping approach}
\label{field}

It can be argued that the article by \citet{VogtVoPeHa1987} (=VPH)
started the field of Doppler mapping as carried out ever since.
After an overview of first efforts to map equivalent widths of
certain metal lines, the authors explain how to recover horizontal
distributions of chemical abundances of various elements. The
ill-posed inversion problem was constrained by the condition that
the resulting map show maximum entropy. The inversion code was
subjected to the famous “Vogtstar” test, based on “V~O~G~T” written
in black letters onto the white surface of a rotating sphere. These
4 letters were reasonably well retrieved by their algorithm, as
were 7 black spots in a further test.

In the 1990s, this approach was developed further by, among others,
\citet{PiskunovWehlau1990} who had a look at the problem of mapping
small and large rings. In contrast to VPH, they used Tikhonov
regularisation \citep{Tikhonov1987} instead of maximum entropy.

Please note that these tests deal with moderately complex patterns of
horizontal inhomogeneities, but they do not include magnetic fields.

\subsection{Constructing an artificial paradise}
\label{arti}

Zeeman Doppler mapping proper really took off with the series of
3 articles in 2002 by O.\,Kochukhov and N.\,Piskunov. In their
second paper, \citet{KochukhovKoPi2002} (=KP) presented numerical
experiments aimed at estimating the accuracy to which the geometry
of the magnetic field could be established by inversion of Stokes
$IQUV$ profiles, and the errors in the horizontal distribution of
chemical abundances derived simultaneously by the same procedure.

The adopted magnetic topology is a combination of a centred dipole
and a linear quadrupole. The Fe abundance distribution is
characterised by
\vspace{-0.14cm}
\begin{itemize}
  \setlength\itemsep{0.15em}
\item either three largish Fe spots (two located at the equator,
  one at a northern latitude, all equally spaced in longitude);
\item or a large single spot centred on the negative magnetic pole;
\item an abundance contrast of 1.5\,dex between the constant
  background and the monolithic spots.
\end{itemize}

These premises ensured reasonably plausible ZDM test results, but
being in no way typical for magnetic CP stars, they cannot be used
to confirm that the Doppler mapping code can actually recover
``fairly complex magnetic and abundance distributions'', as found
for example in HR\,3831 \citep{KochukhovKoDrPiRe2004}, HD\,32633
\citep{SilvesterSiKoWa2015}, or HD\,119419 \citep{RusomarovRuKoLu2018},
but also in 53\,Cam \citep{KochukhovKoBaWaetal2004}.

The authors pointed out that Tikhonov regularisation does not impose
any global constraints on the large scale field structure and that
the resulting magnetic geometry therefore is not necessarily
divergence-free. They professed however to ``{\em believe} that in
all real applications it is possible to find solutions within the
uncertainty of the final magnetic map that would obey integral
constraints imposed by Maxwell’s equations'' [verbatim, our emphasis].
There exists no proof for this supposition, nor was it followed up by
further tests. Nevertheless, non-solenoidal magnetic maps were published
more than a decade after 53\,Cam by \citep{SilvesterSiKoWa2015}.

\subsection{Controversial claims}
\label{claima}

The authors claim that the {\em average} error in the recovered
abundances is $0.04$\,dex and that the reconstruction of the
field modulus is on {\em average} in agreement with the true map
to within 9\%.

Even the shortest glance at Fig.\,5 of \citet{KochukhovKoPi2002}
raises serious doubts as to these estimates, given the poor recovery of
the spots and their surroundings. As can be seen in the difference map
original vs. reconstructed spots (Fig.\ref{orig_maps}), abundances are
found to be locally in error by as much as $\pm\,0.9$\,dex. It is
important to realise that in these tests we are not dealing with errors
in the usual statistical sense involving distributions, Gaussian or
others. They are not caused by photon statistics nor by accumulated
numerical errors in the inversion procedure. Even when exactly the
same physics and the same numerical grids are employed in the forward
synthesis of the Stokes $IQUV$ parameters used for input and in their
inversion, and when the synthesised profiles are fitted to within
$10^{-4}$ or better, we find {\em differences} (in a certain sense
accurate ones), caused by the chosen regularisation functional in the
correct execution of the ZDM  algorithm \citep{StiftLeone2017a}, not
{\em errors}. It follows that one cannot simply exclude ``outliers'' as
done e.g. by \citet{SilvesterSiKoRuWa2017} in the mapping of 49\,Cam. 

The authors maintain that the MDI/ZDM technique provides an
{\em accurate}, self-consistent derivation of magnetic and abundance
maps without positing a specific global magnetic geometry.
Extrapolating straight from the simple magnetic maps discussed above
and the somewhat modest results for the 3 monolithic spots, they
declare that they ``{\em believe} that the code can be successfully
applied to the imaging of global stellar magnetic fields and abundance
distributions of an arbitrary complexity'' [verbatim, our emphasis].

One may legitimately disagree with the view that maps with local
errors/differences of $\pm\,0.9$\,dex can be called {\em accurate}.
Yet even more debatable
is the {\em belief} -- not based on any published comprehensive
test suites or for that matter on any tests at all involving truly
complex magnetic topologies -- that a method working reasonably well for
field geometries defined by just 5 parameters would straightforwardly
lead to unique results for magnetic topologies analysed as a function
of about 400 free parameters, as e.g. in the cases of HD\,32633
\citep{SilvesterSiKoWa2015} and HD\,119419 \citep{RusomarovRuKoLu2018}.

Once ZDM analyses had suggested chemical distributions of ``stunning
complexity'' and magnetic geometries that could not be modelled by
dipole plus quadrupole fields, extensive realistic tests should have
followed the 2002 paper. The fact that the assessment of the credibility
of almost the totality of published ZDM results and claims has been
based exclusively on a {\em belief} enunciated back in 2002 is quite
disconcerting.

\subsection{More idealisations, more claims}
\label{more}

There is just one other paper presenting numerical tests of Doppler
mapping \citep{Kochukhov2017} (=K17) which however only deals with
abundances, not with magnetic fields. This time the star features
4~spots, placed at $-30\degr$, $0\degr$, $+30\degr$ and $+60\degr$
latitude, and equidistantly spaced in longitude (as back in 2002).
The spots are no longer assumed monolithic but display a central part
with a constant [Fe/H] = -2.5 and a smooth run from the inner radius
of $15\degr$ to the outer radius of $30\degr$ where it merges into the
constant background with [Fe/H] = -4.0. Regrettably, none of the spots
in the test coincide with magnetic poles, a constellation found in a
number of stars, e.g. in HR\,3831 \citep{KochukhovKoDrPiRe2004}(Ca, Ba),
$\alpha^2$\,CVn \citep{SilvesterWaKoLaBa2008}(Mg) or HD\,125248
\citep{RusomarovRuKoRyIl2016}(Cr). The test is not applicable to
the assessment of the pole-centred Li, O, Mn, Fe, and Nd spots in
HD\,3980, star credited with a field of 7\,kG polar strength
\citep{NesvacilNeLuShetal2012}.

There are conspicuous anomalies in the published test results. K17 
clearly states that his test is based on a grid of 1876 approximately
equal-area pixels in 38 latitude belts. This is in conflict with the
spherical plots in his Fig.\,3a which he labels as ``true abundance
maps'' and with subsequent figures. The input maps for the synthesis
of the $IQUV$ spectral line profiles to be inverted are unpixeled
and in fact correspond exactly to the idealised spots at infinite
resolution. The nature of the recovered maps is more difficult to
ascertain: not only is there no trace whatsoever of the 1876 pixel
inversion, but scans along the x- or y-coordinates reveal remarkably
smooth RGB profiles with about 5 times the number of surface elements
(and corresponding colours) compared to a 1876 pixel grid.

This immediately raises the question of how to interpret the inverted
maps in Fig.\,3b,c and later; they cannot be the product of heavy
smoothing of a relatively coarse grid by an unspecified algorithm.
Regardless of the method used to create these unpixeled maps from the
1876 pixel ZDM results, differences with respect to the original maps
are inevitably ironed out to some degree. The published maps thus no
longer exhibit the true differences/errors. In the present case the
situation is further complicated by the ambiguous nature of the
``original''. Is it the 1876 pixel map or the idealised
infinite-resolution map? We cannot choose between these scenarios
without more information or detailed modelling. 

\begin{figure}
  \centering
  \includegraphics[width=90mm, angle=270]{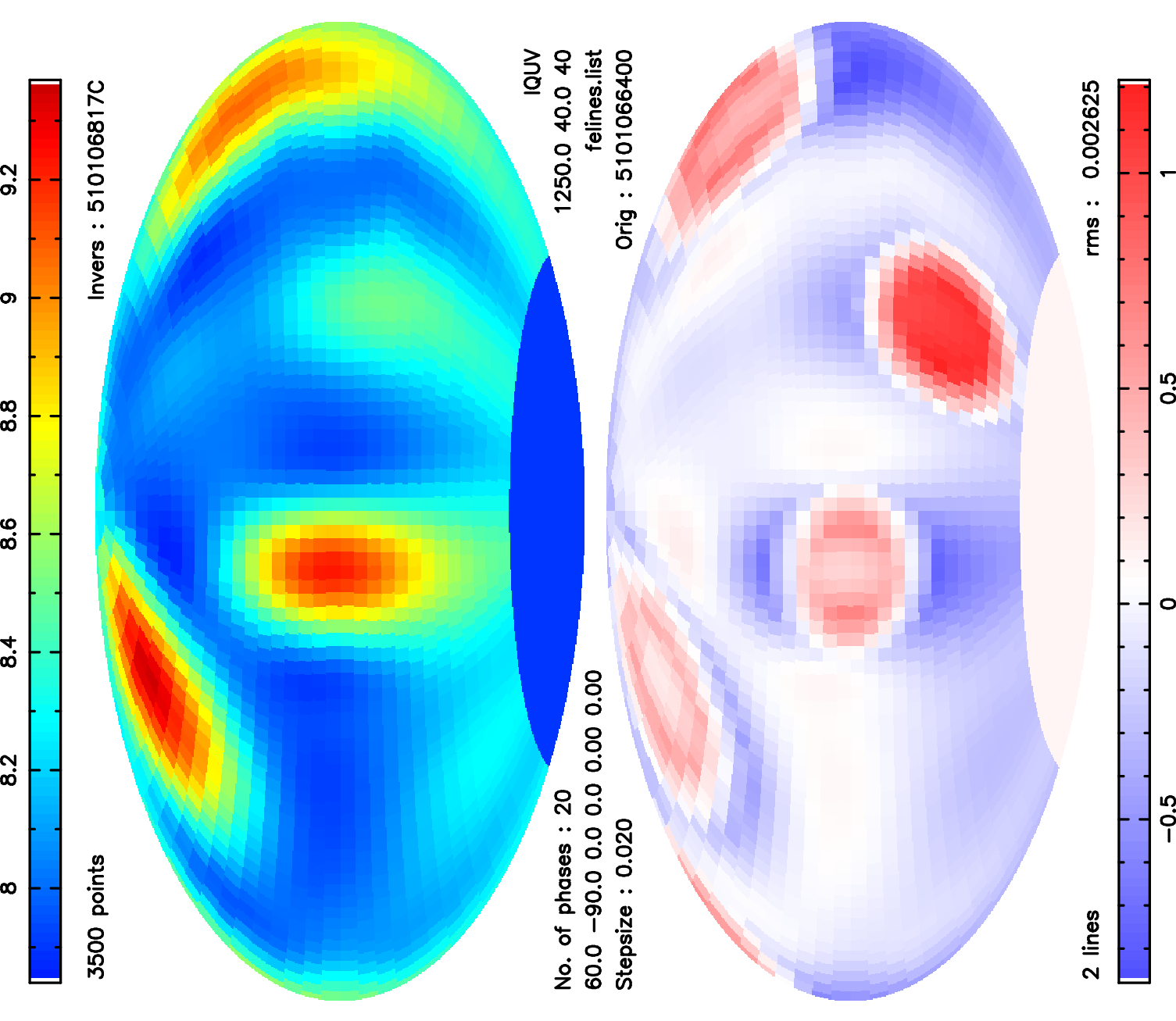}
  \caption{
    Hammer projection of the recovered abundance map for the K17 4-spot
    test case, based on all 4 Stokes parameters at 20 phases, 20\,m{\AA}
    spectral resolution and 2034 pixel spatial grid.
    {\bf Top:} Unsmoothed ZDM output.
    {\bf Bottom:} Difference map original minus inverted.
}
\label{2017_maps}
\end{figure}

There is an additional unknown to the appreciation of the difference
maps: what has happened to abundance value outliers that according
to \citet{SilvesterSiKoRuWa2017} normally comprise 2–12\% of all the
surface elements. Have they been excluded prior to smoothing? And how
in this context do we have to interpret the approach applied by K17?
In view of the overabundance regions (the 4 spots) covering 27\% of
the stellar surface, he considered the 75th percentile of the unsigned
difference map as a relevant measure of the maximum inversion errors.
What happened to the 470 pixels exceeding the maximum error, have they
been excluded, replaced or interpolated?

In order to elucidate these problems, we decided to revive the ZDM
code presented in \citep{StiftLeone2017a} and to simulate the K17
test for the 2 Fe lines. Adopting a 2034 pixel grid in 40 latitude
belts, the same spot geometries and the same dipole magnetic field
as in K17, we synthesised $IQUV$ profiles for various numbers of
phases up to 20 and at spectral resolutions of 20\,m{\AA}, 25\,m{\AA},
and 50\,m{\AA}. We did the same for a 10310 pixel grid in 90 latitude
belts. Fig.\ref{2017_maps} shows the result of the inversion for the
K17 fwo-line case. In sharp contrast to the K17 plots, the pixeled
nature of the inverted map is clearly visible, and so is the pixeled
difference map between the true input map and the true, unsmoothed
recovered map. The $IQUV$ spectral line profiles at 3500 wavelength 
points (altogether 14000 values) have been fitted to better than
$3\,10^{-3}$.

\begin{figure}
  \centering
  \includegraphics[width=90mm, angle=270]{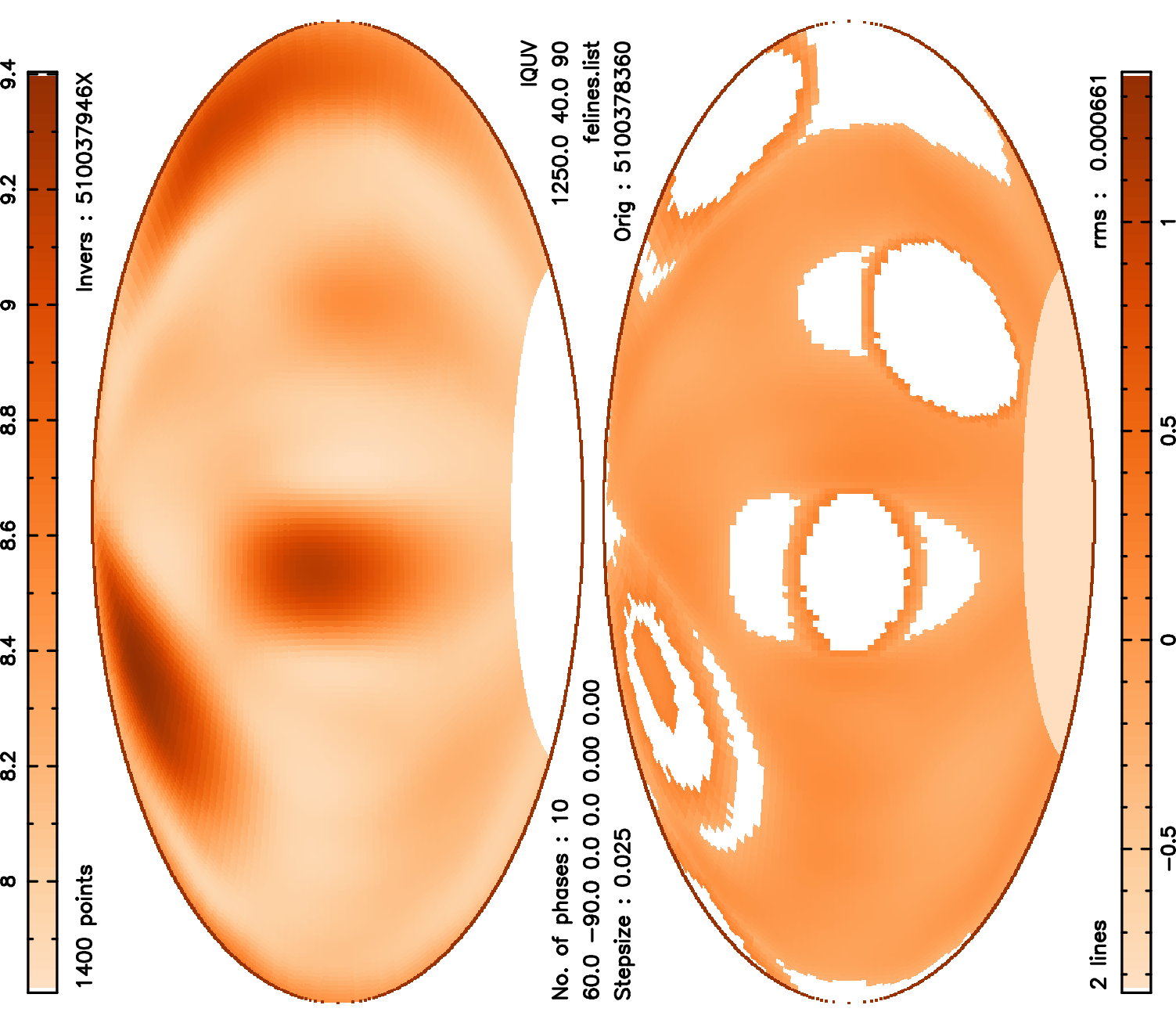}
  \caption{
    Same as before, but now based on all 4 Stokes parameters at
    10 phases, 25\,m{\AA} spectral resolution and 10310 pixel spatial grid.
    {\bf Top:} Unsmoothed ZDM output.
    {\bf Bottom:} Difference map original minus inverted; the excluded
    pixels are displayed in white.
}
\label{2017_sel_maps}
\end{figure}

Our results are in serious disagreement with the ``maximum error''
estimates given by K17, viz. 0.13 to 0.15\,dex, once the uppermost
quartile of the ``errors'' is discarded. As pointed out above, in
this kind of test we are dealing with {\em differences} that 
Fig.\ref{2017_maps} shows to be almost 8 times larger than claimed
by K17. The abundance of large parts of the southern spot is
underestimated by 1\,dex and more, whereas the spot at $+30\degr$
is stretched in the N-S direction beyond the equator, with the
northern part quite underabundant, the southern part overabundant
by up to 0.8\,dex. Let us emphasise that these maximum differences
change very little ($\pm 0.1$\,dex) with number of phases, spectral
resolution and size of spatial grid.

Certainly, the poor performance of the K17 DM inversion regarding
in particular the spot at $-30\degr$ latitude is not propitious
to the credibility of the abundance maps published for several CP
stars with inclination near $60\degr$. In HD\,3831, Na, Si, Ca,
Mn, Fe, and Y spots are located well to the south of the equator
\citep{KochukhovKoDrPiRe2004}. In HD\,133880, the most conspicuous
spots of Mg, Si, Ti, Cr, and Fe are likewise found in the partly
hidden southern hemisphere \citep{KochukhovKoSiBaLaWa2017}. Ca, Cr,
and Eu in 49\,Cam, fare similarly, but this time north of the
equator \citep{SilvesterSiKoRuWa2017}.

We also would like to add that the decision by K17 to throw out the
pixels belonging to the uppermost ``error'' quartile is lacking a
statistical foundation, making it impossible to assess the effects
of this choice on the results. Taking our inversions, we would have
to exclude those 470 pixels from the maps in Fig.\ref{2017_maps} that
differ by more than 0.305\,dex from the original. We illustrate the
outcome with a 10310 pixel inversion (Fig.\ref{2017_sel_maps}) --
5600 wavelength points in $IQUV$ fitted to $6.6\,10{-4}$ -- where the
exclusion of 2577 pixels differing by more than 0.235\,dex is seen to
severely affect all spots (and their surroundings), rendering them
virtually invisible.

\subsection{Outdated and skewed numerical tests}
\label{skewed}

As discussed above, the two numerical tests presented so far are 
unsatisfactory as both the adopted magnetic field geometries and
the horizontal abundance inhomogeneities are overly simple. For the
2002 paper this can be considered acceptable since at that time people
had only vague ideas on what kind of magnetic field geometry could be
expected and horizontal chemical structures would look like. Not so
for the K17 paper which in no way reflects the wealth and variety of
maps published between 2002 and 2017. Please keep in mind that the
entirety of ZDM data constitutes a fairly homogeneous set, having been
analysed exclusively with the codes of O.\,Kochukhov; publications
have without exception been written by himself or closely supervised.

Take a handful of examples. For 53\,Cam \citet{KochukhovKoBaWaetal2004}
found a magnetic field topology considerably more complex than any low
order multipolar expansion. \citet{KochukhovKoWa2010} and later
\citet{SilvesterSiKoWa2014a} showed that the observed Stokes profiles
of $\alpha^2$\,CVn could not be fitted with a dipole plus quadrupole
geometry. This was repeated by \citep{SilvesterSiKoWa2015} for
HD\,32633 and by \citep{RusomarovRuKoRyIl2016} for HD\,125248.

\citet{KochukhovKoDrPiRe2004} claimed a ``stunning complexity'' of the
horizontal Fe abundance pattern in HR\,3831. The surface distribution
of Ca is fairly complex too, without large areas of overabundance or
underabundance. In the case of HD\,125248 \citep{RusomarovRuKoRyIl2016}.
the very strong surface abundance inhomogeneities also do not bear any
resemblance to the 4-spot pattern adopted by K17.

For HR\,3831, \citet{KochukhovKoDrPiRe2004} have assumed a dipole field
of 2.5\,kG polar strength. Small Li and Ba spots are concentrated in
areas at or very close to the magnetic poles. In the case of HD\,3980
\citep{NesvacilNeLuShetal2012} with an estimated polar field strength
of 7\,kG, spots of Mn, Pr, O, Li, Nd have been found at the poles.
Tests with large spots on the other hand are exclusively located away
from the poles and are therefore misleading.

\section{Assessing published, ZDM based maps}
\label{maps}

An interesting dichotomy emerges. On one side, two highly idealised
numerical tests \citep{KochukhovKoPi2002, Kochukhov2017} seem to
indicate ``maximum errors'' in the recovery of horizontal abundance 
distributions of just 0.13 to 0.15\,dex. On the other side, what we
prefer to call ``maximum differences'' can reach values in excess
of 1\,dex as demonstrated by \citet{StiftLeone2017a} and again in
the present study, this time in a rerun of the 4-spot test of K17. 
Is it possible to obtain supplementary information that would allow
us to choose between these conflicting estimates and provide a sound
base for the realistic assessment of the validity and accuracy of
abundance and magnetic field maps?

It is always challenging to compare results arrived at with different
codes, but most fortunately there is a solution that entirely avoids
these problems: stick solely to inversions effected with the
{\sc invers} family of codes. ZDM analyses carried out over the years
with largely identical methods and codes by O.~Kochukhov and a small
group of researchers under his supervision will guarantee a high
degree of internal consistency. Providentially for our endeavour, a
number of magnetic CP stars have been analysed several times. Separate
maps of the same star have been derived from data sets that can be
identical, just slightly differing or completely distinct; published
over the years or decades, they ensure meaningful comparisons that are
not influenced by varying approaches to ZDM or by different codes.

Be aware that we do not look at the plausibility of the physical and
mathematical assumptions underlying the ZDM procedure. As mentioned in
section\ref{intro}, \citet{StiftLeone2022} have already convincingly
shown that a number of ZDM analyses are violating the laws of
magneto-hydrodynamics.
Here we are exclusively interested in the interagreement between the
ZDM results obtained and presented by O.\,Kochuknov and his group.
Ideally, differences between abundance maps of the same star should
not be markedly in excess of the values given by K17, i.e. 0.15\,dex.
For the magnetic field one would then expect errors of roughly
10-20\%. For early critical reviews of ZDM analyses we refer to
\cite{StiftStLeCo2012} and to \cite{StiftLeone2017b, StiftLeone2017a}.

\begin{figure}
\includegraphics[height=90mm, angle=270]{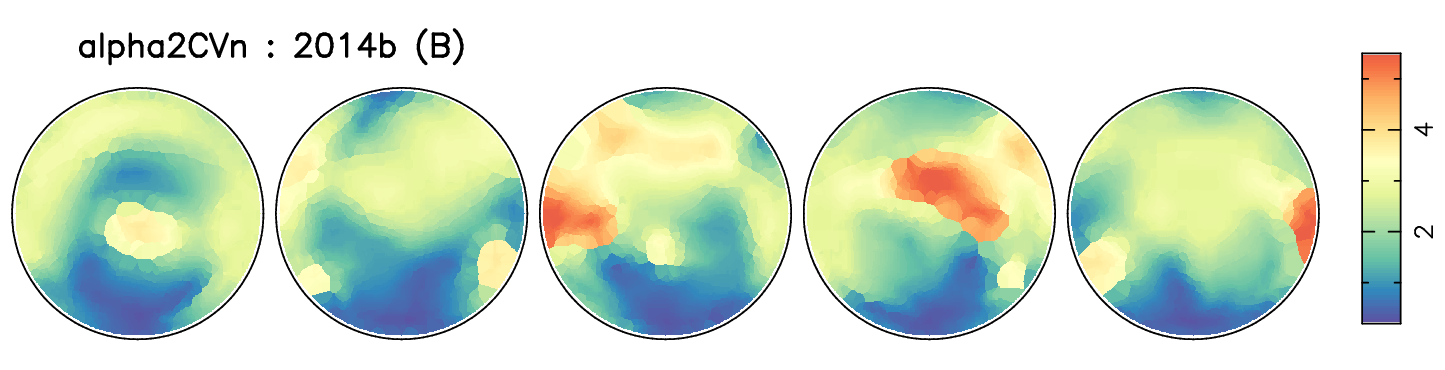}
\includegraphics[height=90mm, angle=270]{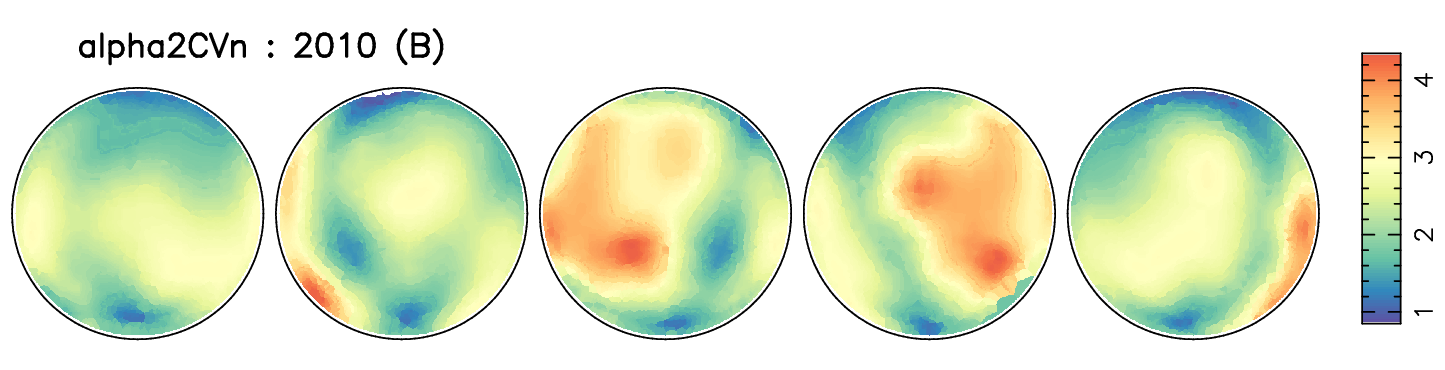}
\includegraphics[height=90mm, angle=270]{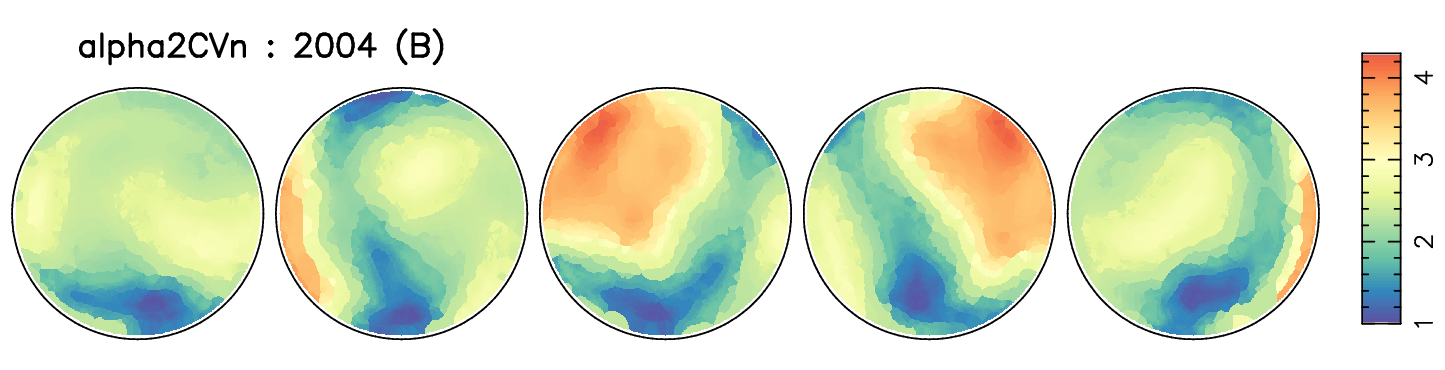}
\includegraphics[height=90mm, angle=270]{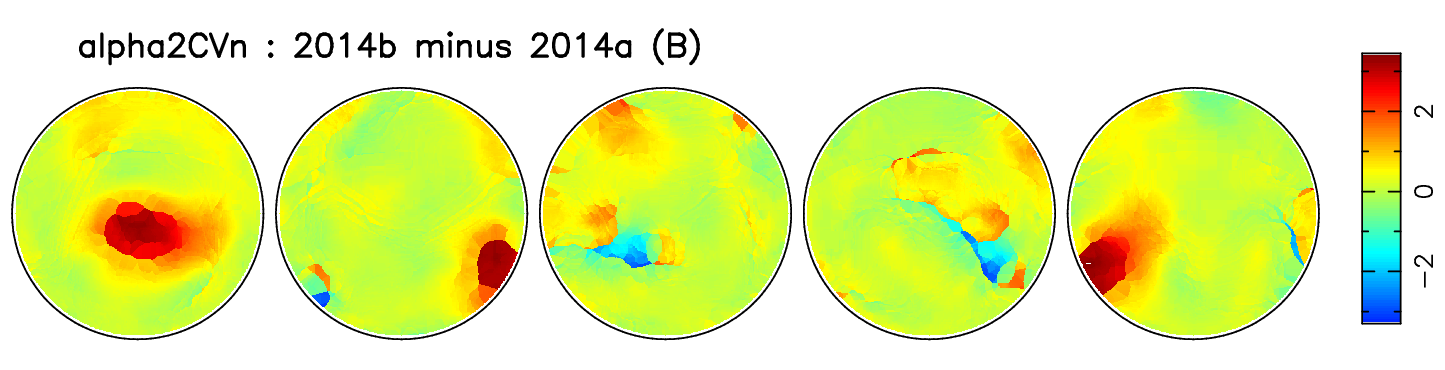}
\includegraphics[height=90mm, angle=270]{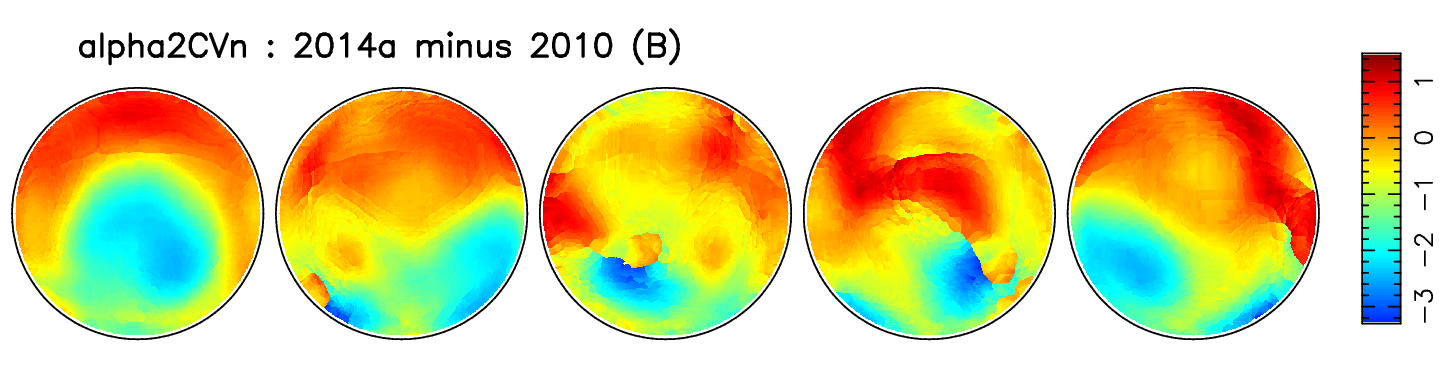}
\includegraphics[height=90mm, angle=270]{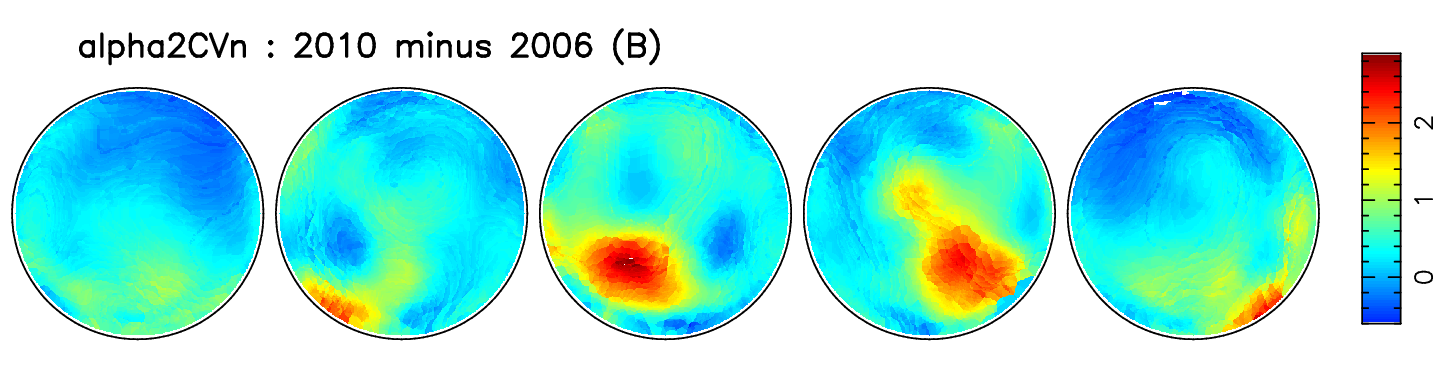}
\caption{
  ({\bf top 3}) $\alpha^2$\,CVn : Absolute magnetic field strengths at 5
  equidistant phases, recovered from 3 articles. ({\bf bottom 3}) Differences
  in field strengths at 5 equidistant phases for combinations between 5 maps.
}
\label{Alpha_B_m_diff}
\end{figure}

\section{Data extraction and its caveats}
\label{data}

In contrast to the situation 10-20 years ago, high quality colour plots in
online publications now make it possible to numerically recover detailed
ZDM results on abundances and magnetic fields. Code of just a handful of
lines written in a popular interpreted computing language helps to obtain
data from which -- after a few coordinate transformations -- detailed maps
can be established (see e.g. \citealt{StiftLeone2022}).

There are some stumbling blocks that render the process of data recovery
challenging, if not sometimes impossible:
\begin{itemize}
\item In the worst case \citep{SilvesterWaKoLaBa2008}, there is no colour
  wedge to convert colours to abundances or field strengths.
\item Image compression artefacts in plots of poor quality -- as encountered
  in \citet{NesvacilNeLuShetal2012}) -- can always seriously complicate
  data recovery, but in the particular case of the first ZDM study of
  $\alpha^2$\,CVn \citep{KochukhovKoPiIlTu2002}, all our efforts have failed.
\item Maps are not always accompanied by vital data such as the inclination
  $i$ of the rotational axis. Only after more than a decade, 49\,Cam had
  a value of $i$ published for the first time, disappointingly without any
  mention of the previously used, differing value(s).
\item Isolines in plots of magnetic field components (radial, meridional,
  azimuthal) can display strange behaviour at the borders as in Fig.\,6 of
  \citet{SilvesterSiKoWa2015}. Unfortunately we cannot simply ignore this
  and extract the underlying data (which are needed for example in the
  modelling of stratified abundances for the star), because
  discontinuities occur in all 3 components with small gaps of unspecified
  size around phase 0.0/1.0.
\item Slide\,19, entitled ``Stability of Ap-star spots'', in a lecture
  given by O.\,Kochukhov 2014 in Besan\c{c}on,
  {\footnotesize (lesia.obspm.fr/MagMas/cartography2014/kochukhov\_lecture.pdf)}
  displays Si spots on 56\,Ari that seem to move in longitude. In the same
  lecture, on slide\,39, we find the original 2002 test case exhibiting the
  same behaviour. Numerous other instances turn up in articles published
  between 2002 and 2014, among them \citet{KochukhovKoDrRe2003},
  \citet{Piskunov2008}, and \citet{LuftingerLuKoRyetal2010b}.
\end{itemize}

\section{A detailed discussion of 5 stars}
\label{stars}

In the following we shall discuss 5 stars for which multiple abundance and/or
magnetic maps could be found in journals, proceedings, and web pages. There
are altogether 12 different maps of field strength and 20 different maps of
Cr, Fe, and Nd available for our study. Very few of the contributions to
proceedings have ever been cited in subsequent papers, refereed or not, so
the following discussion will give, for the first time, an overview of these
maps, their variety, their disparateness and their chronology.

\subsection{$\alpha^2$\,CVn (HD\,112413)}
\label{alpha}

$\alpha^2$\,CVn  is the star with the most published maps. The first magnetic
map based on full Stokes $IQUV$ spectral line profiles is due to
\citet{Kochukhov2004IAUS224}, followed shortly after by \citet{Kochukhov2006}.
Different results were presented by \citet{KochukhovKoWa2010}, superseded
by 2 papers a couple of years later \citep{SilvesterSiKoWa2014a,
SilvesterSiKoWa2014b}. Chromium and iron maps can be found in
\citet{KochukhovKoWa2010}, \citet{SilvesterSiKoWa2014b}. and in an undated
plot (file name alpha2CVn\_abn.jpg) that seems to correspond largely to the
2002 B/W plot, available until several years ago at an Uppsala University
homepage. This plot is also found in a 2014 Uppsala lecture note on inverse
problems.

\begin{figure}
\includegraphics[height=90mm, angle=270]{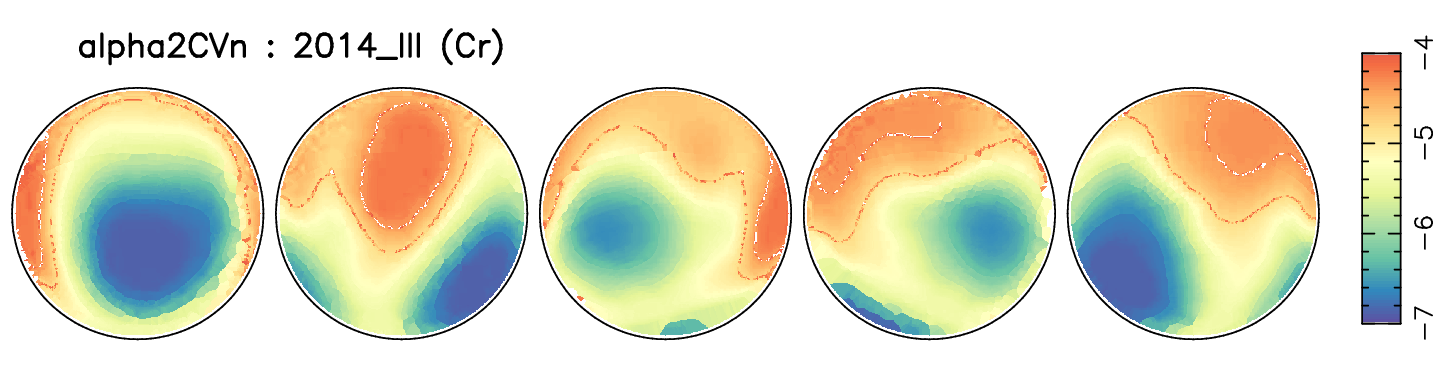}
\includegraphics[height=90mm, angle=270]{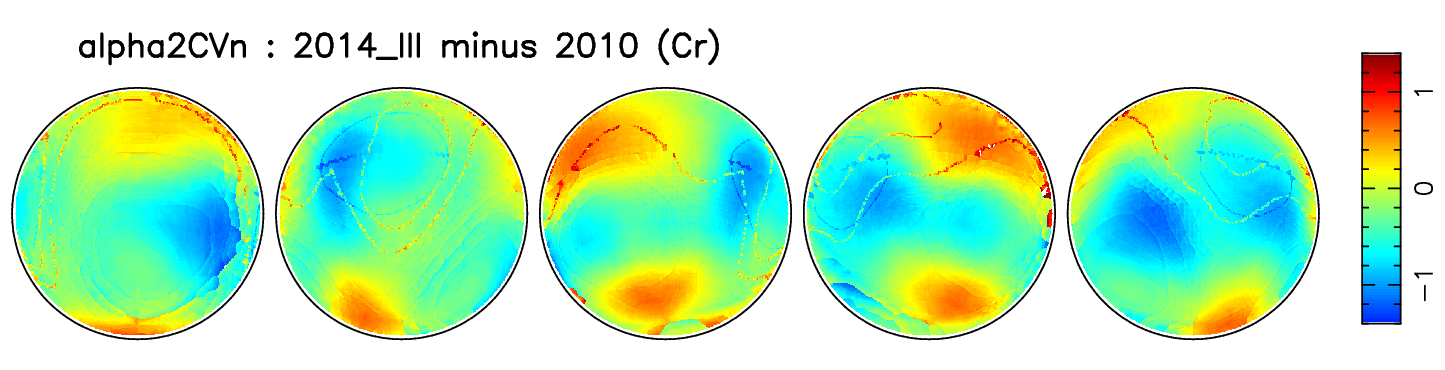}
\includegraphics[height=90mm, angle=270]{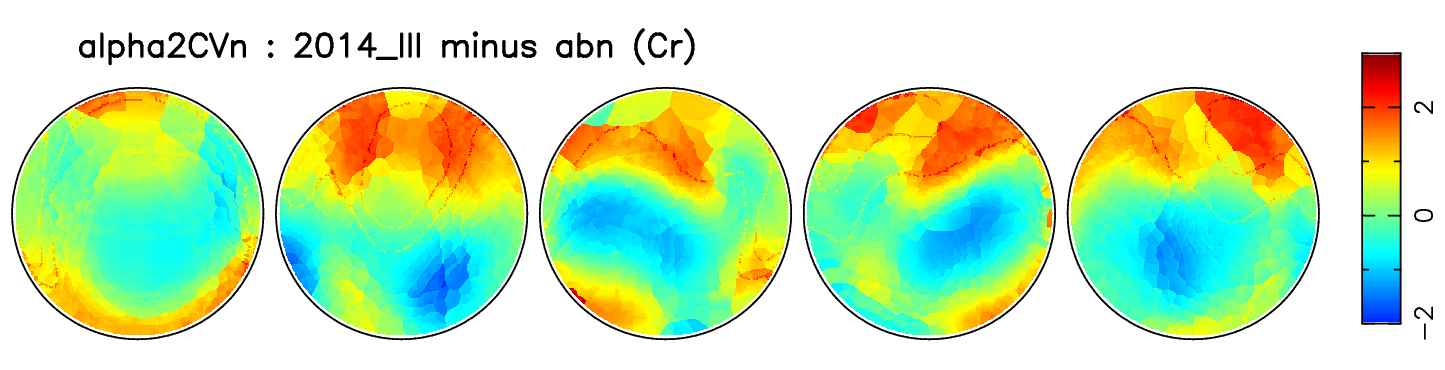}
\includegraphics[height=90mm, angle=270]{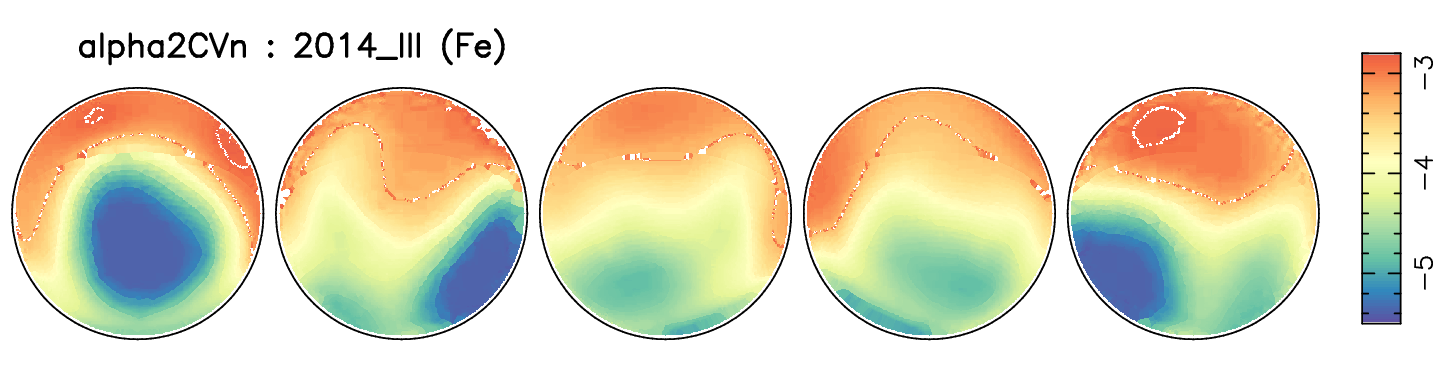}
\includegraphics[height=90mm, angle=270]{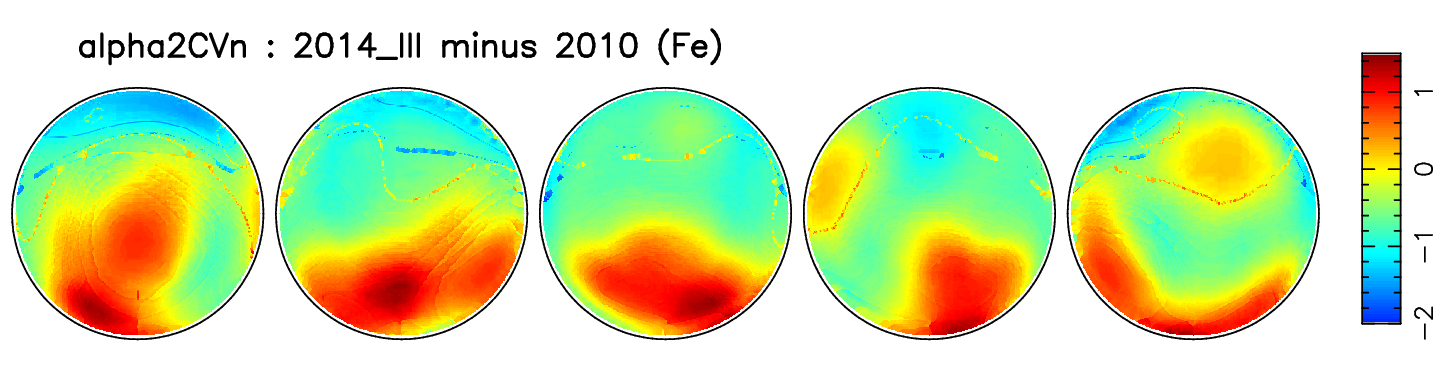}
\includegraphics[height=90mm, angle=270]{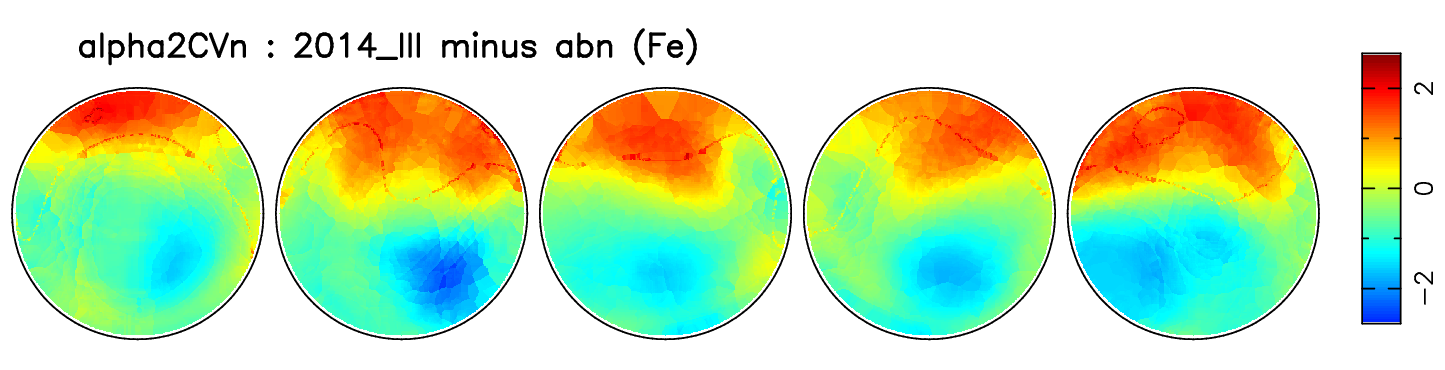}
\caption{
  ({\bf top 3}) $\alpha^2$\,CVn : Cr abundance at 5 equidistant phases and
  abundance differences for combinations between 3 maps.
  ({\bf bottom 3}) Same as before, but for Fe abundances.
}
\label{Alpha_CrFe_m_diff}
\end{figure}

The striking dissimilarities between the field maps stand out visually and
do not require sophisticated tools to quantify (Fig.\ref{Alpha_B_m_diff}).
Still, only differential maps fully reveal the differences which in the
case 2014b minus 2014a run from $-3.3$\,kG to $+3.4$\,kG. The 2014a maps
are based on 5 iron and 3 chromium lines in Stokes $IQUV$, to which 2
oxygen and chlorine lines in Stokes $IV$ were added for the 2014b maps.
It is somewhat surprising that this modest increase in the number of lines
would lead to such a substantial revision of the magnetic map, but similar
discrepancies have already previously occurred when the addition of 2 Cr
lines to the 2010 analysis (based on 5 Fe lines and 1 Cr line) led to the
2014a map that differs from 2010 by $-3.3$\,kG to $+1.5$\,kG.

It is true that the resolution (R=35.000) of the MuSiCoS spectra underlying
the 2010 maps is no match to the resolution of ESPaDOnS and Narval (R=65.000).
However, as \citet{SilvesterSiKoWa2014a} have pointed out, there is good
agreement between the two data sets, and the overall phase sampling is quite
similar. It seemed to them therefore ``reasonable'' to compare field maps
derived with the same atomic line list. The huge differences in field
strengths were not pointed out at that time.

Abundances hardly fare any better than field strengths
(Fig.\ref{Alpha_CrFe_m_diff}). The 2014b chromium map featuring a $3.0$\,dex
contrast differs by $-1.4$ to $+1.4$\,dex from the undated map and by $-2.0$
to $+3.0$\,dex from the 2010 map. Fe displays a contrast of $2.7$\,dex, and
differences are large again, reaching $-2.0$ to $+1.5$\,dex in the first
case, $-2.7$ to $+2.7$\,dex in the second.

These results contradicting the conclusions by K17 that there is little
difference between maps based on 5 and on 2 lines, we may therefore 
legitimately question the alleged quality of the recovered maps. Discussion
of the other 4 stars will further nourish our doubts.

We note that none of the results published before 2010 has ever been
cited in later articles, only the 2010 paper being mentioned in 2014.
More than 10 years later, the mechanisms leading to conflicting field
and abundance maps still remain unexplored. Additional evidence for
multiple solutions will be presented below.

\subsection{49\,Cam (HD\,62140)}
\label{49Cam}

In all, 2 different Fe maps and 4 different magnetic field maps of 49\,Cam
can be found in three papers and one slide published between 2007 and 2017.
\citet{Silvester2007magneticdopplerimagingap} offered the very first iron
and field maps, followed a few years later \citep{SilvesterWaKoLaBa2011}
by a field map based on 14 spectra  (compared to 8 spectra back in 2007).
Finally, \citet{SilvesterSiKoRuWa2017} came up with new Fe and field maps,
derived from 19 spectra. With just 5 spectra added, one would expect little
difference between the 2017 and the 2011 field maps, but see below. There
exists one more field map, contained in a presentation by J.\,Silvester at
the STARS2016 Conference
(https://www.star.uclan.ac.uk/stars2016/slides/Silvester.pdf).
Given the time stamp, this map ought to be based on the same 19 spectra
as the 2017 map.

\begin{figure}
\includegraphics[height=90mm, angle=270]{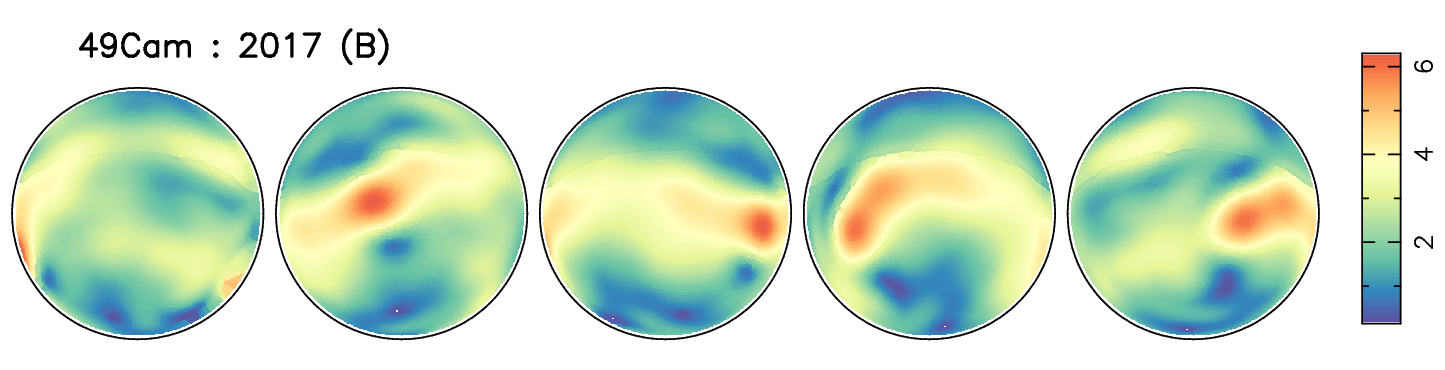}
\includegraphics[height=90mm, angle=270]{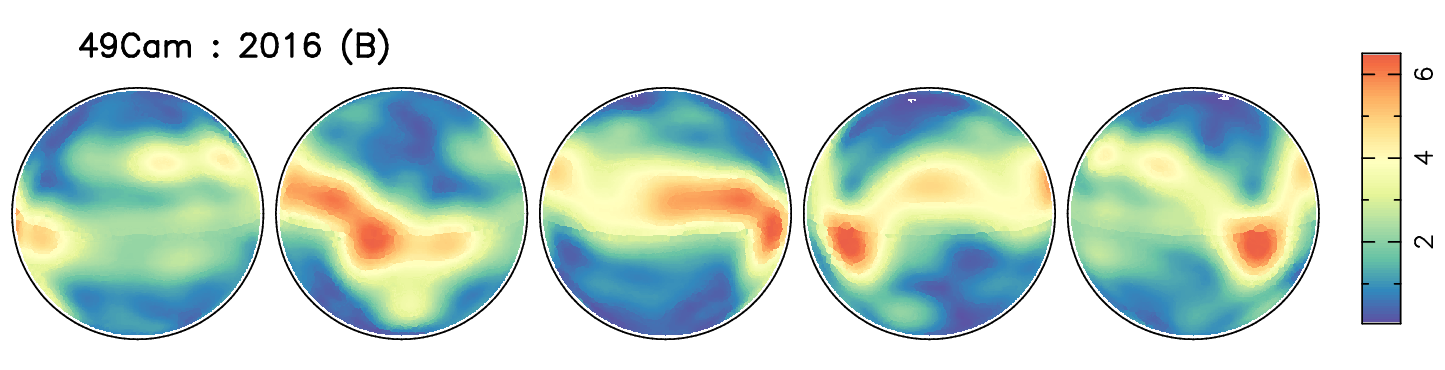}
\includegraphics[height=90mm, angle=270]{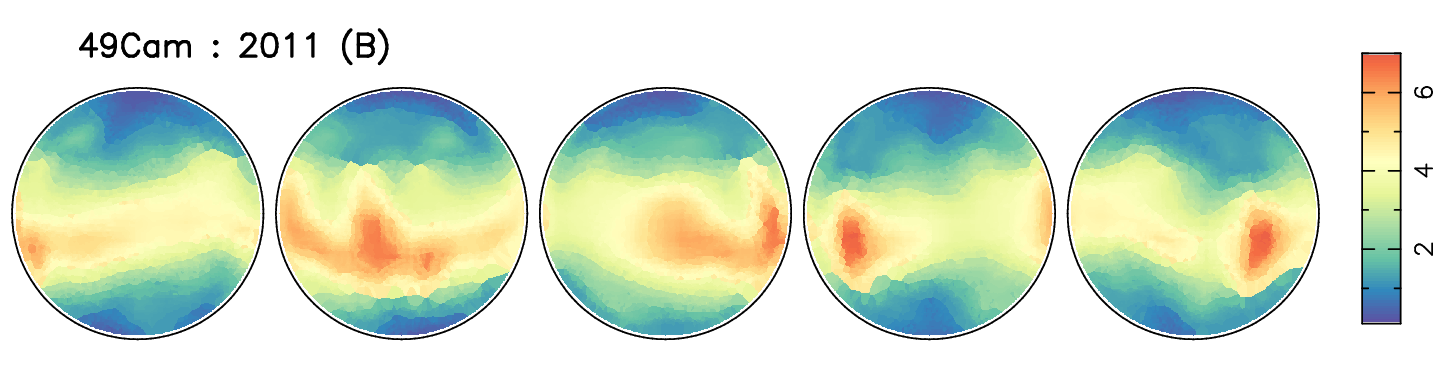}
\includegraphics[height=90mm, angle=270]{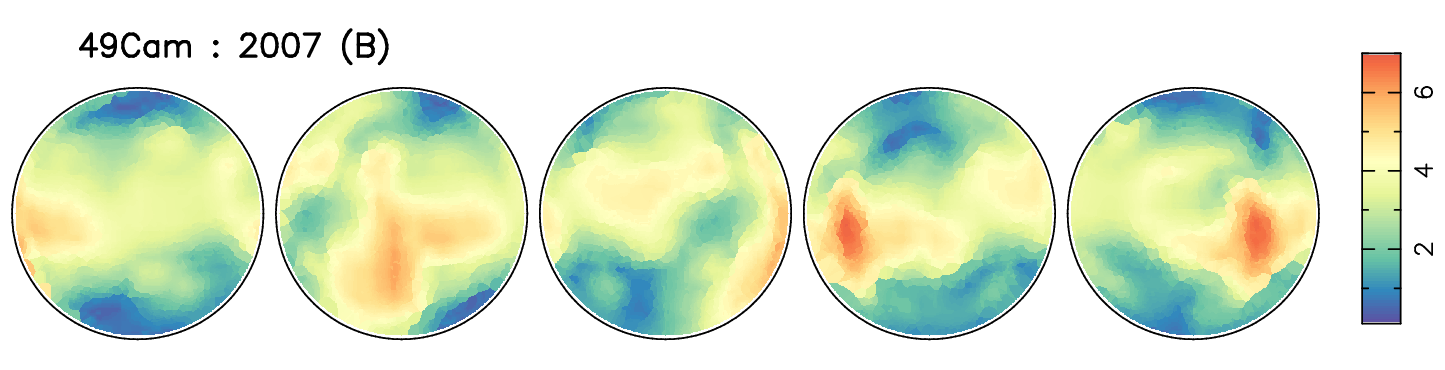}
\includegraphics[height=90mm, angle=270]{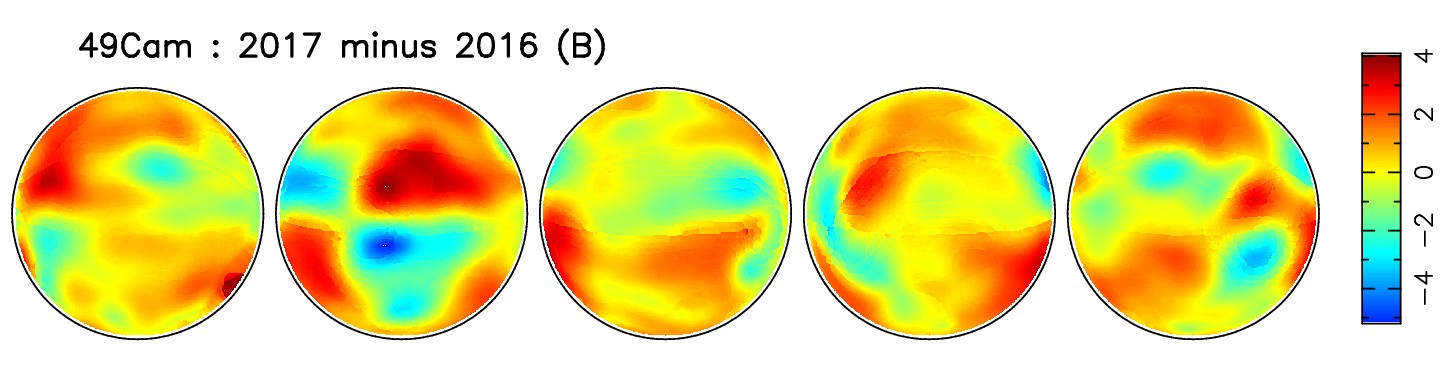}
\includegraphics[height=90mm, angle=270]{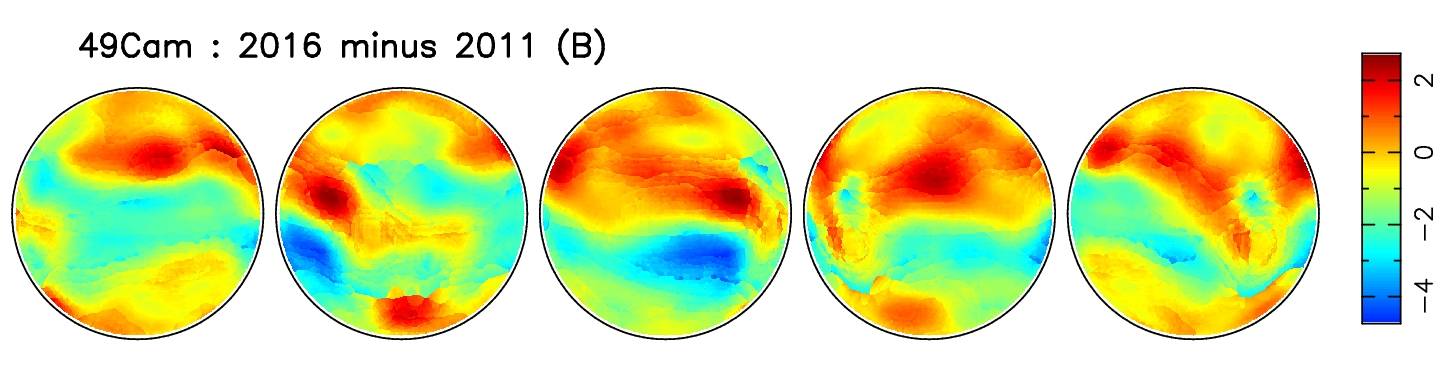}
\includegraphics[height=90mm, angle=270]{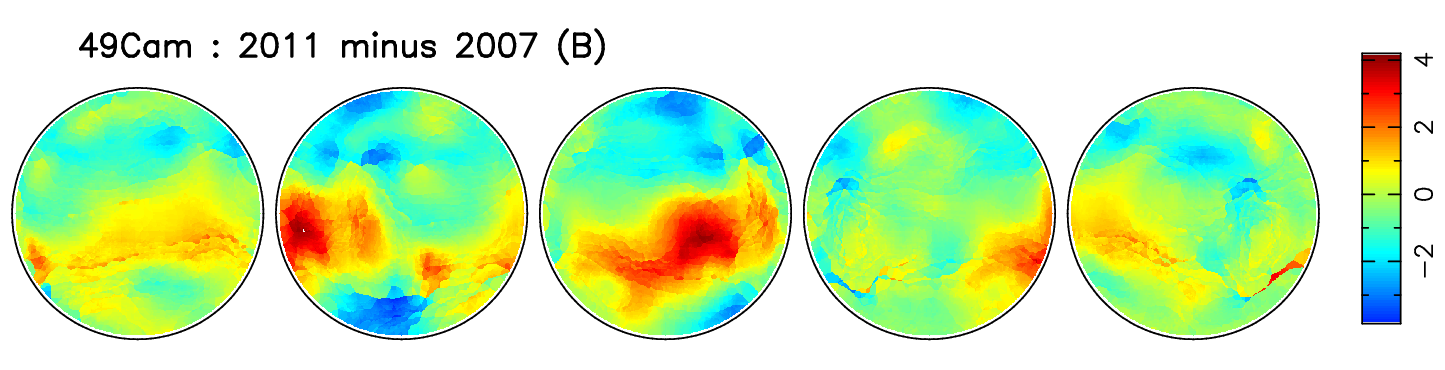}
\caption{
  ({\bf top 4}) 49\,Cam : Absolute magnetic field strengths at 5
  equidistant phases. ({\bf bottom 3}) Differences in field strengths
  at 5 equidistant phases for combinations between the 4 maps.
}
\label{49Cam_B_m_diff}
\end{figure}

The magnetic map in the arXiv preprint and the largely identical
corresponding final article \citep{SilvesterWaKoLaBa2008} differs
substantially from the map presented in the contribution to the
meeting ``Astronomical Polarimetry 2008'' \citep{SilvesterWaKoLaBa2011}.
where no mention is made of the 2007/2008 map.

\begin{figure}
\includegraphics[height=90mm, angle=270]{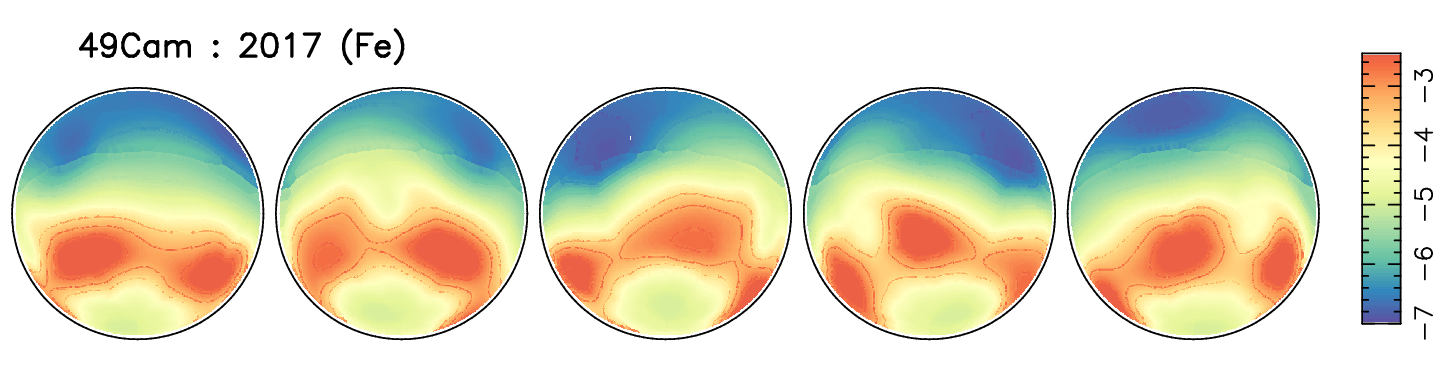}
\includegraphics[height=90mm, angle=270]{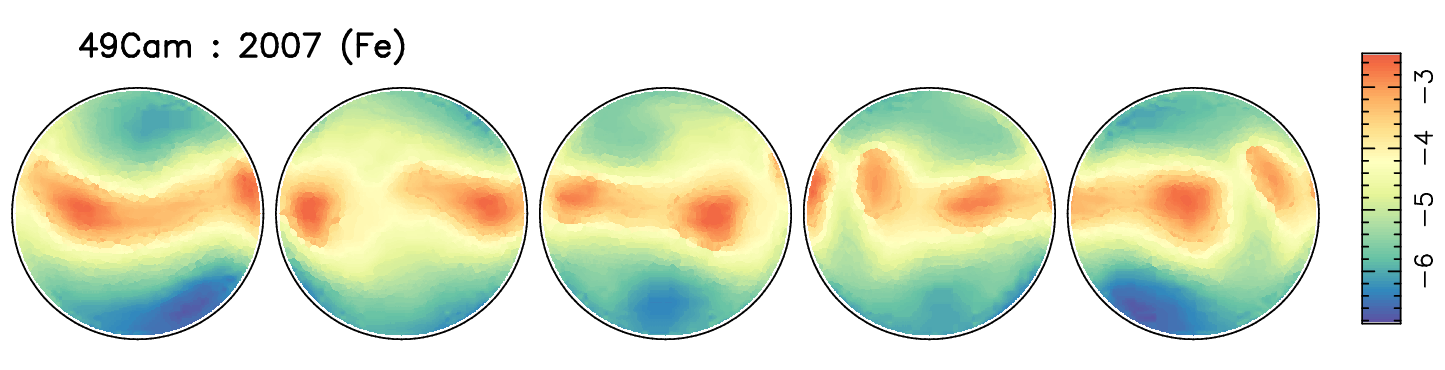}
\includegraphics[height=90mm, angle=270]{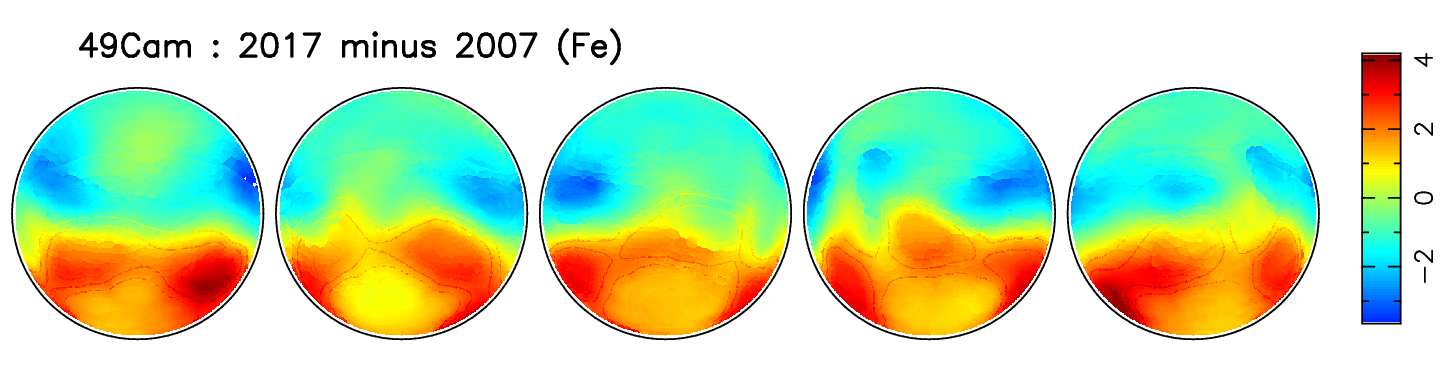}
\caption{
  ({\bf top 2}) 49\,Cam : Fe abundances at 5 equidistant phases.\\
  ({\bf bottom}) Fe abundance differences between the 2 maps.
}
\label{49Cam_Fe_m_diff}
\end{figure}

The reasons for the discrepancies between the 2007/2008 and the 2011
results -- both called ``(very) preliminary'' by the authors -- have
never been discussed. In 2017, no mention at all can be found of the
2007 and 2011 papers, a fact that also holds true for the 2016 slide.
Whereas a mere 8 spectra might fall short of providing the necessary
input for reliable abundance or field maps, 14 spectra should be
sufficient, and 19 spectra can be expected to plainly confirm the
2011 findings.

The abundance contrast in the 2 iron maps is almost identical at
$\approx 4,5$\,dex. but local differences 2017 minus 2007 are found to
range between $-3.6$\,dex and $+4.2$\,dex (Fig.\ref{49Cam_Fe_m_diff}).
Similarly, all 4 field maps (Fig.\ref{49Cam_B_m_diff}).
agree on an approximate contrast of $6$ to $7\,$kG. Yet, concerning the
magnetic geometry, there is no agreement at all: differences in field
strength 2017 minus 2011 cover a staggering $-5.6$\,kG to $+3.4$\,kG
(remember the 19 vs. 14 spectra !). 2016 minus 2011 is only slightly
less spectacular with $-4.7$\,kG and $+2.7$\,kG. For 2011 minus 2007
we find extrema of $-3.8$\,kG and $+4.2$\,kG.

This brings us to a curious feature in the remarkable evolution of the
magnetic maps. Having adopted -- as probably in all previous analyses
-- an inclination of $85\degr$ of the rotational axis for his 2016
slide, \citet{SilvesterSiKoRuWa2017} suddenly changed this to $120\degr$.
Of course adjustments to the inclination are not unusual in ZDM: over
the years, the inclination of $\alpha^2$\,CVn has changed by $10\degr$,
by $17\degr$ for HD\,24712. One could also argue with the 2002 and 2017
numerical tests that showed magnetic inversions to be relatively
insensitive to moderate uncertainties ($\approx 10-20\degr$) in the
inclination angle. Here however the adjustment in the angle reaches 
perplexing  $25$ degrees, and it turns out that both 2017 field and Fe
maps bear precious little resemblance to their predecessors from 2007
to 2016.

We think that the fascinating history of the 49\,Cam ZDM maps should
not slip into oblivion because of lack of citations in the definitive
study. The discrepancies are sufficiently troubling to
warrant an in-depth discussion. As stated in the previous subsection,
a convincing case can be made again for the existence of multiple
solutions to ZDM inversions.

\begin{figure}
\includegraphics[height=90mm, angle=270]{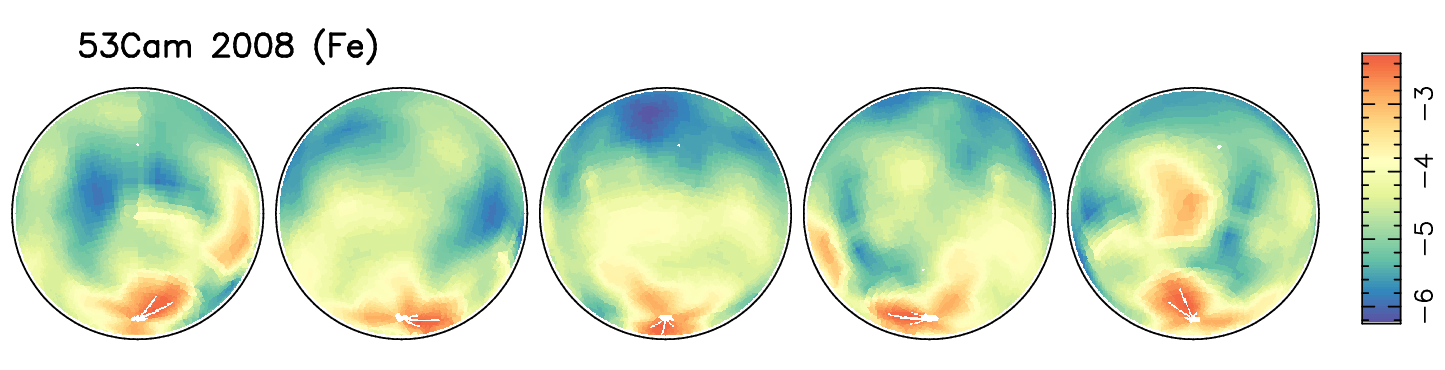}
\includegraphics[height=90mm, angle=270]{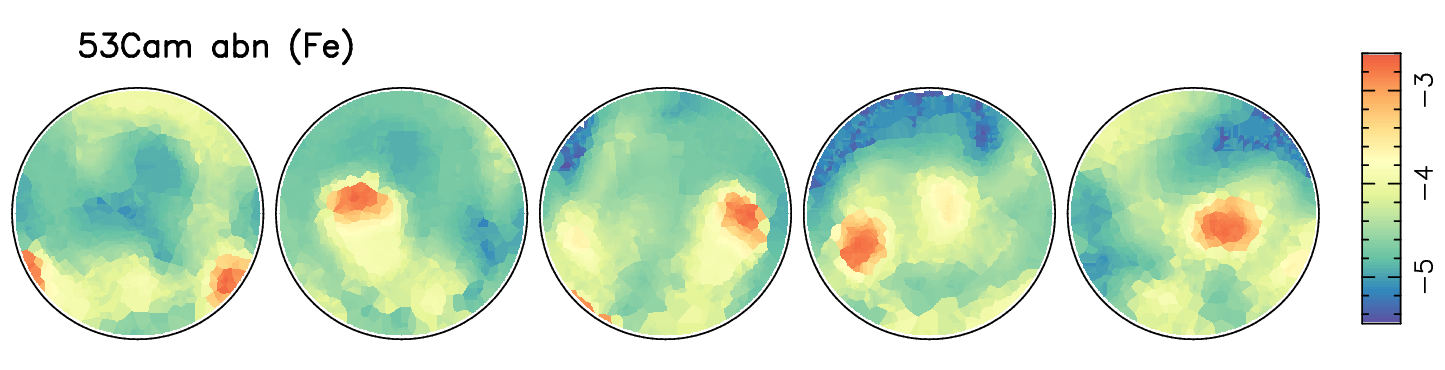}
\includegraphics[height=90mm, angle=270]{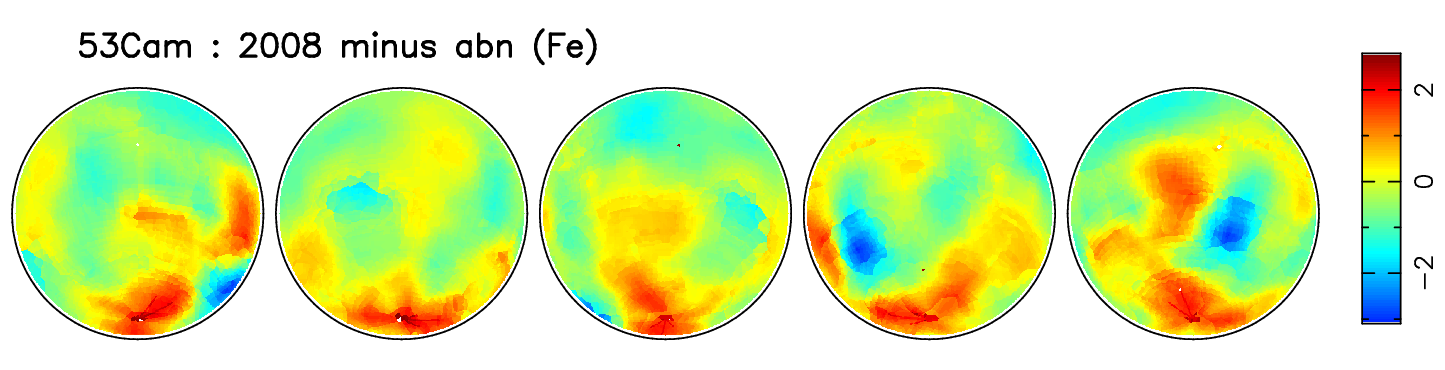}
\caption{
  ({\bf top 2}) 53\,Cam : Fe abundances at 5 equidistant phases.\\
  ({\bf bottom}) Fe abundance differences between the 2 maps.
}
\label{53Cam_Fe_m_diff}
\end{figure}

\subsection{53\,Cam (HD\,65339)}
\label{53Cam}

53\,Cam was the first CP star to be observed and analysed in Stokes
$IQUV$ \citep{KochukhovKoBaWaetal2004}. The recovered magnetic topology
exhibits field strengths of up to $27$\,kG. It is neither divergence-free,
nor does it obey the force-free condition. It would thus seem that this
star is of no use for our present study. However, there are 2 maps of
iron abundances available, based on the same set of observations. In
addition to the B/W plots in the 2004 paper as republished in colour by
\citet{Piskunov2008}, there existed an undated plot (53Cam\_abn.jpg), 
similarly to alpha2CVn\_abn.jpg available on an Uppsala University
homepage.

\begin{figure}{b}
\includegraphics[height=90mm, angle=270]{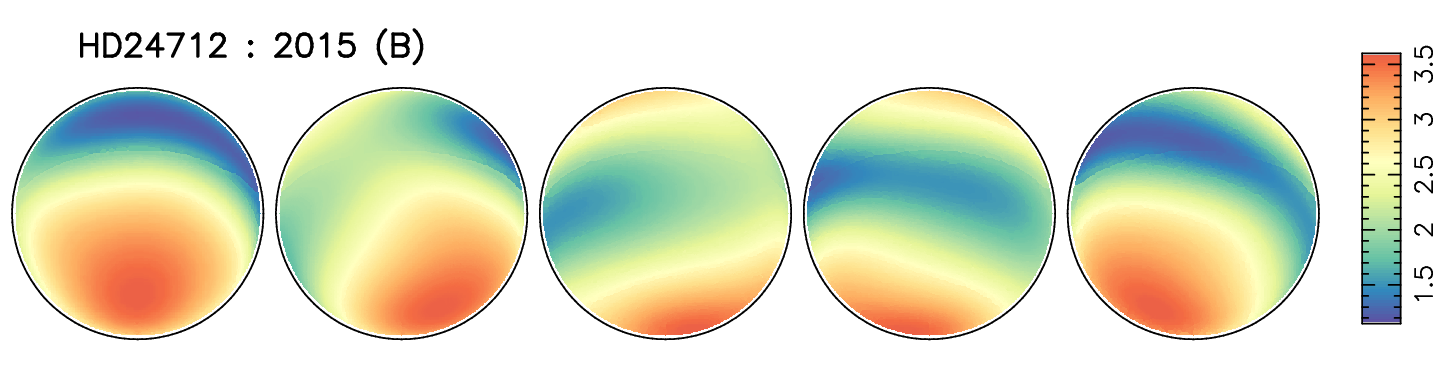}
\includegraphics[height=90mm, angle=270]{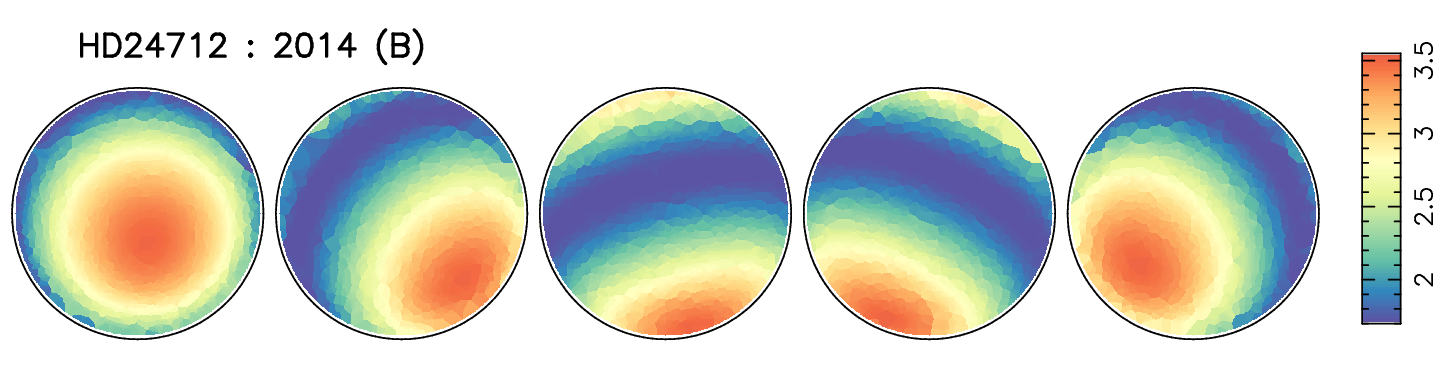}
\includegraphics[height=90mm, angle=270]{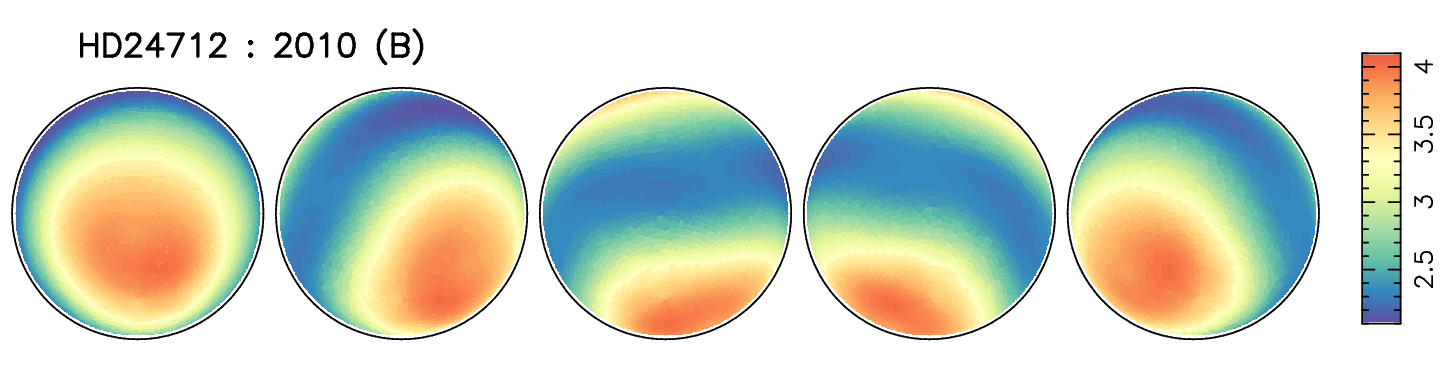}
\caption{
  (HD\,24712 : Absolute magnetic field strengths at 5 equidistant phases.\\
}
\label{HD242712_B_m}
\end{figure}

\begin{figure}
\includegraphics[height=90mm, angle=270]{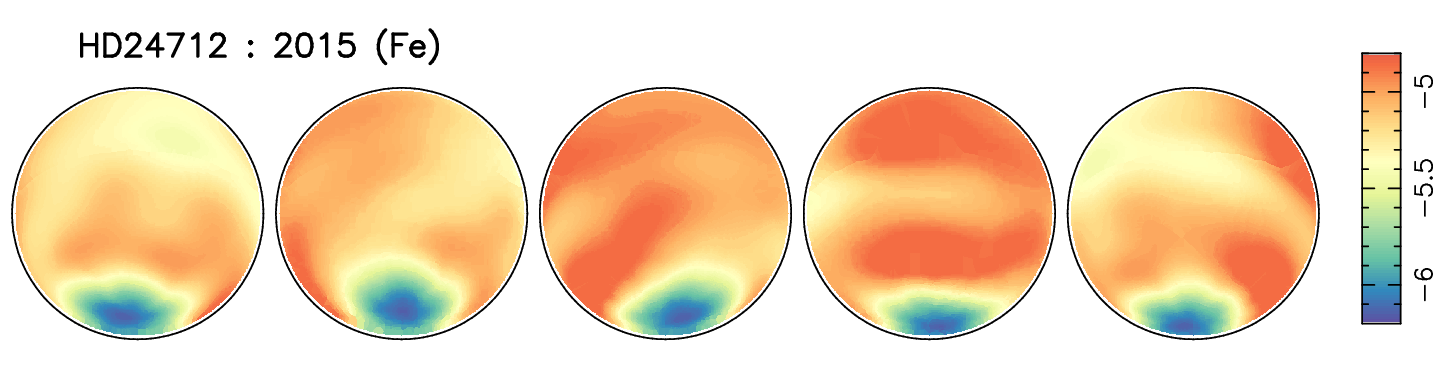}
\includegraphics[height=90mm, angle=270]{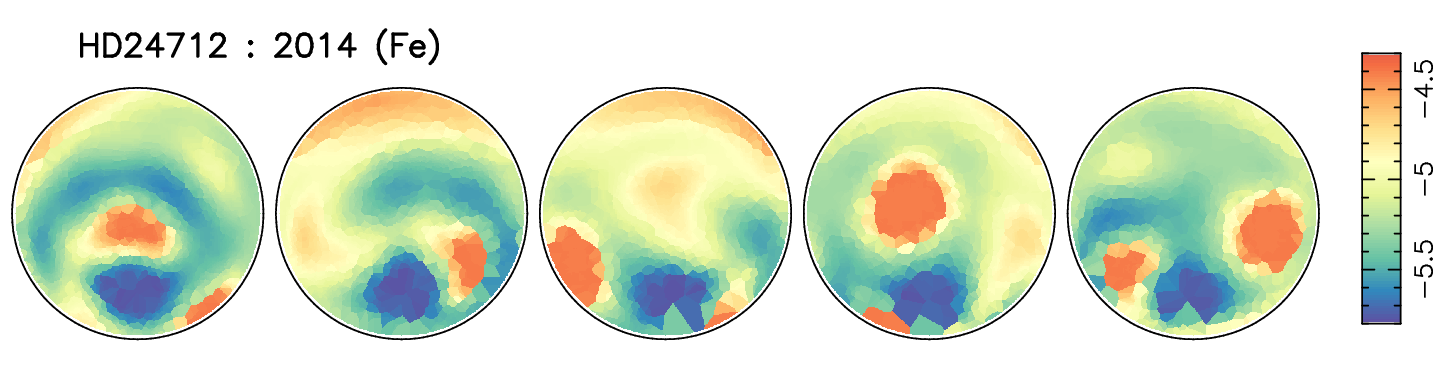}
\includegraphics[height=90mm, angle=270]{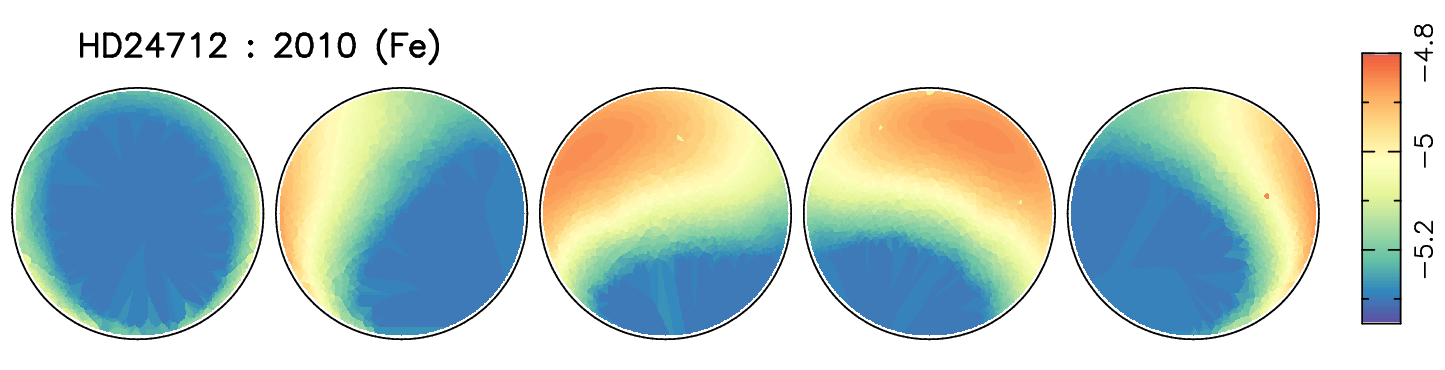}
\includegraphics[height=90mm, angle=270]{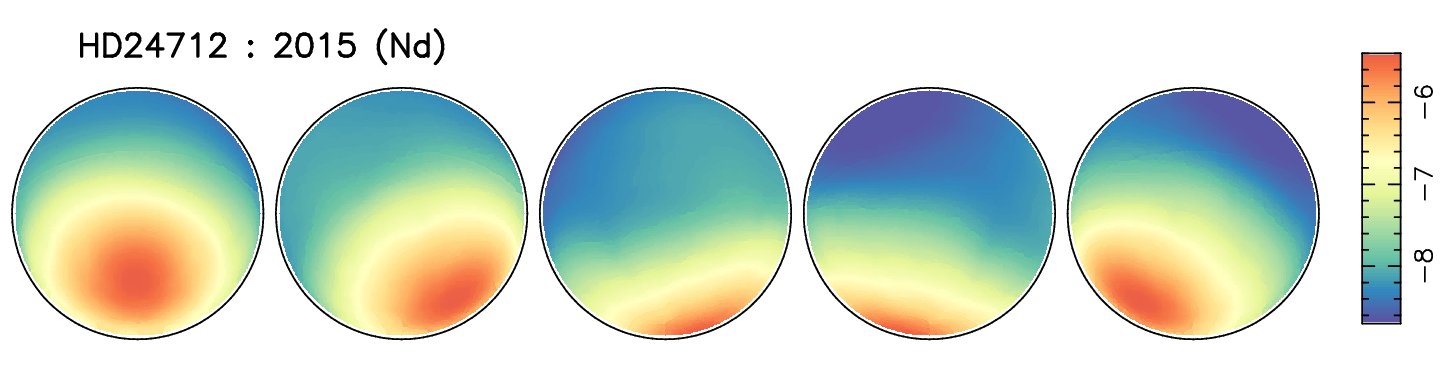}
\includegraphics[height=90mm, angle=270]{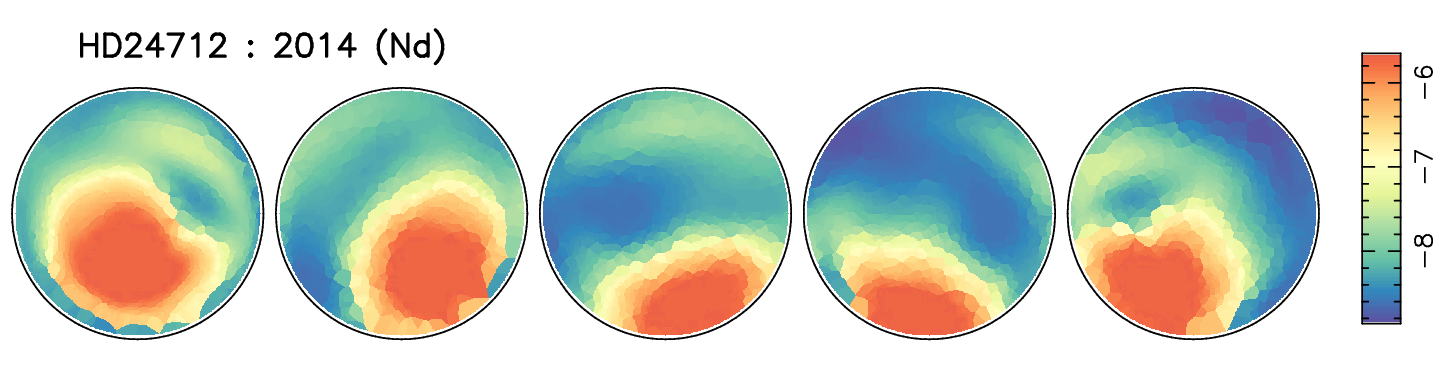}
\includegraphics[height=90mm, angle=270]{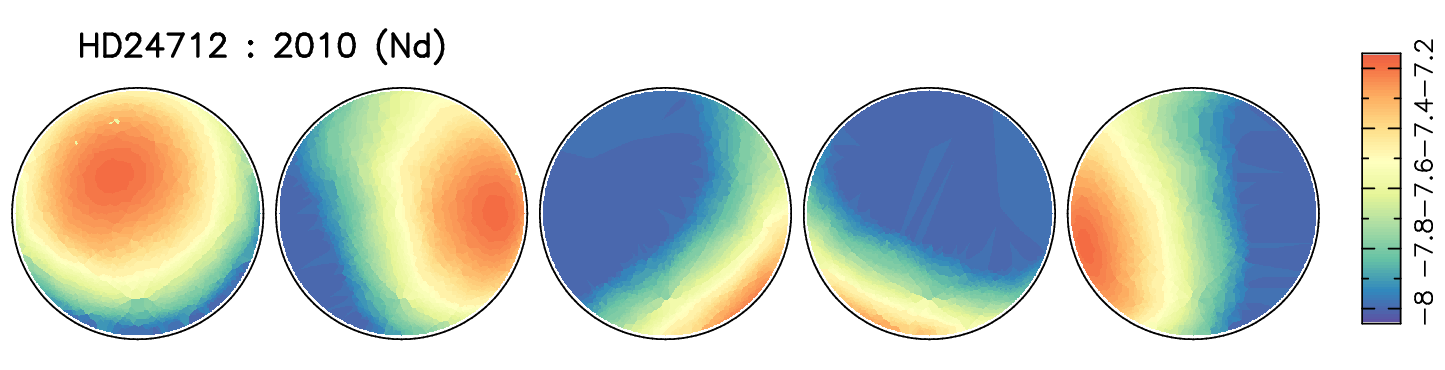}
\caption{
  ({\bf top 3}) HD\,24712 : Fe abundances at 5 equidistant phases.\\
  ({\bf bottom 3}) Nd abundances at 5 equidistant phases.\\
}
\label{HD242712_FeNd_m}
\end{figure}

Here too, the maps prove discordant, both with respect to the location
of the small spots and (to a lesser extent) to the contrast which
amounts to $4.0$\,dex (2008) and $2.9$ (undated/abn)
respectively. The differences 2010 minus abn range from $-3.1$\,dex to
$+2.8$\,dex. Fig.\ref{53Cam_Fe_m_diff} clearly reveals the size of the
discrepancies. Once again, the optimistic error estimates given by K17
turn out unrealistic when it comes to real-life ZDM.

We note that \citet{KochukhovKoWa2010} mention an ''updated'' magnetic
map of 53\,Cam, produced in the same way as for $\alpha^2$\,CVn, that
allegedly does not differ {\em appreciably} from the {\em average}
field topology inferred back in 2004. The vagueness of this wording
does not help in assessing the actual differences, and so is their
Fig.\,12 which mainly reveals a fairly strong (but unphysical) toroidal
field. The relative strengths of the poloidal field components differ
manifestly from those in Fig.\,11 of the 2004 paper, and so do the
toroidal components. We strongly encourage the authors to publish the
``updated'' magnetic and abundance maps which should clarify the true
nature, structure and extent of the differences 2010 vs. 2004.

\subsection{HD\,24712}
\label{HD24712}

HD\,24712 is a roAp star with a fairly low rotational velocity of
$v\,\sin i = 5.6\,{\rm km\,s}^{-1}$ which is not conducive to good
spatial resolution with respect to latitude. Relying on Stokes
$IV$ only and adopting an inclination angle of $i = 137\degr$,
\citet{LuftingerLuKoRyetal2010b} determined a dipolar magnetic
geometry and derived abundance maps for a number of chemical
elements, with emphasis on Fe and Nd. Iron was found to concentrate
near the equator over a considerable interval in longitude, with
neodymium almost at antiphase near $-30\degr$ south. Inversion of
high-quality (R=114.000) HARPSpol spectra in Stokes $IQUV$ however
led \citet{Rusomarovetal2014} to results quite at variance with
these 2010 maps. Fe now featured 2 high contrast spots of moderate
size and an underabundant south polar region, whereas Nd could
hardly be considered in antiphase. The position of the Nd spot
had shifted by at least $30\degr$ in latitude as can be seen in
Fig.\ref{HD242712_FeNd_m}.

\begin{figure}
\includegraphics[height=90mm, angle=270]{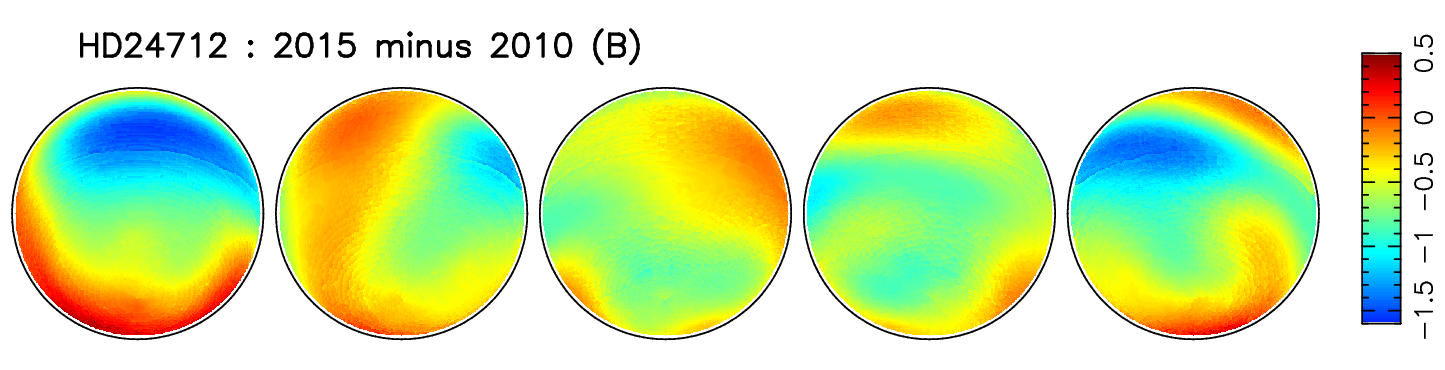}
\includegraphics[height=90mm, angle=270]{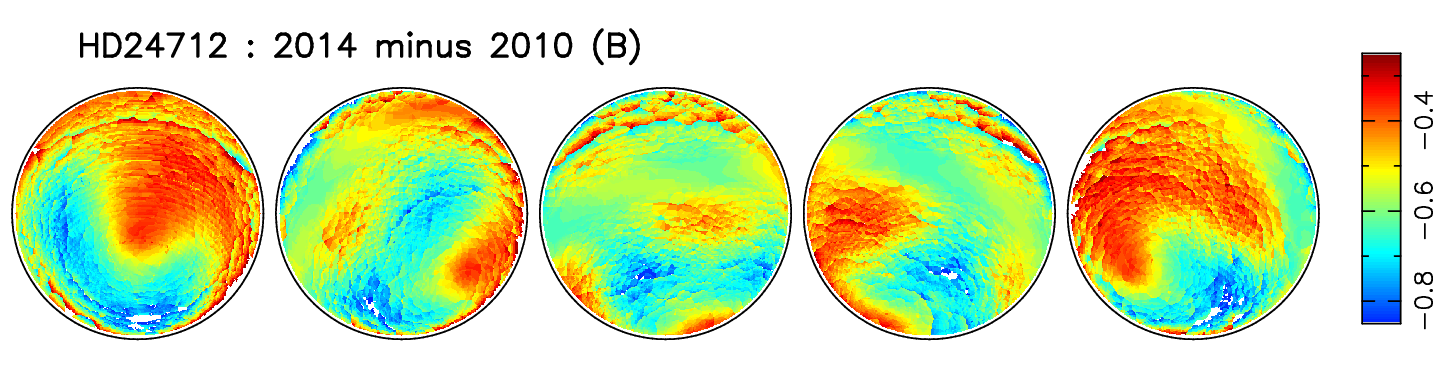}
\includegraphics[height=90mm, angle=270]{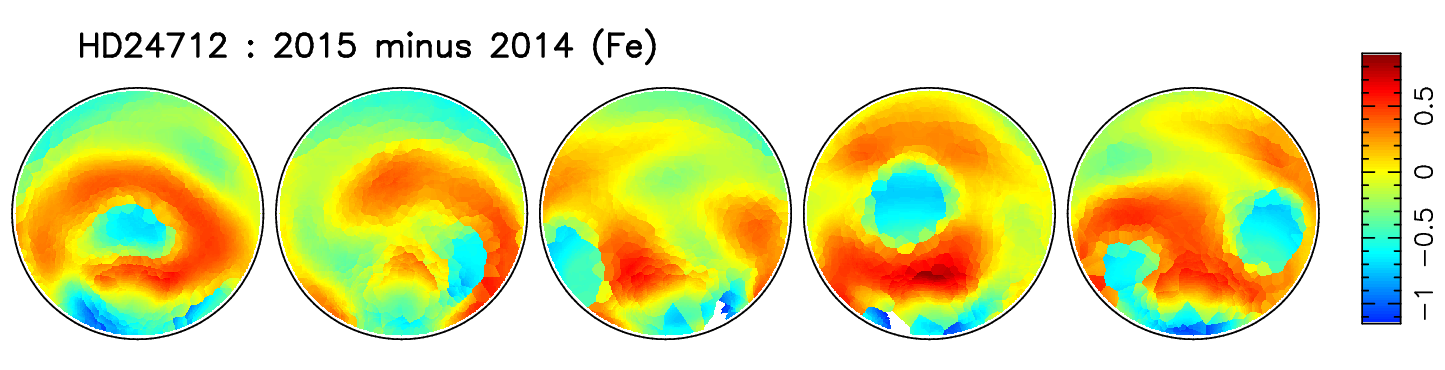}
\includegraphics[height=90mm, angle=270]{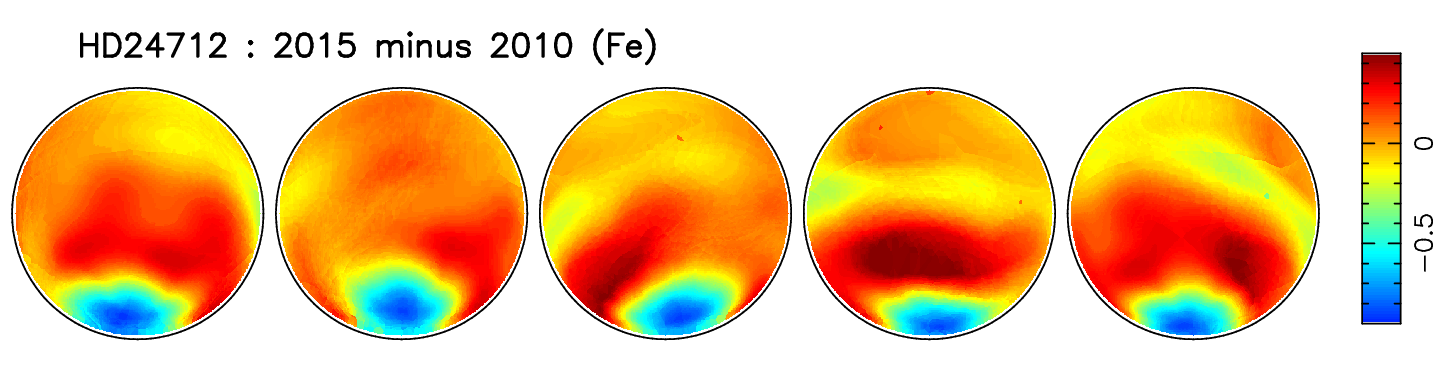}
\includegraphics[height=90mm, angle=270]{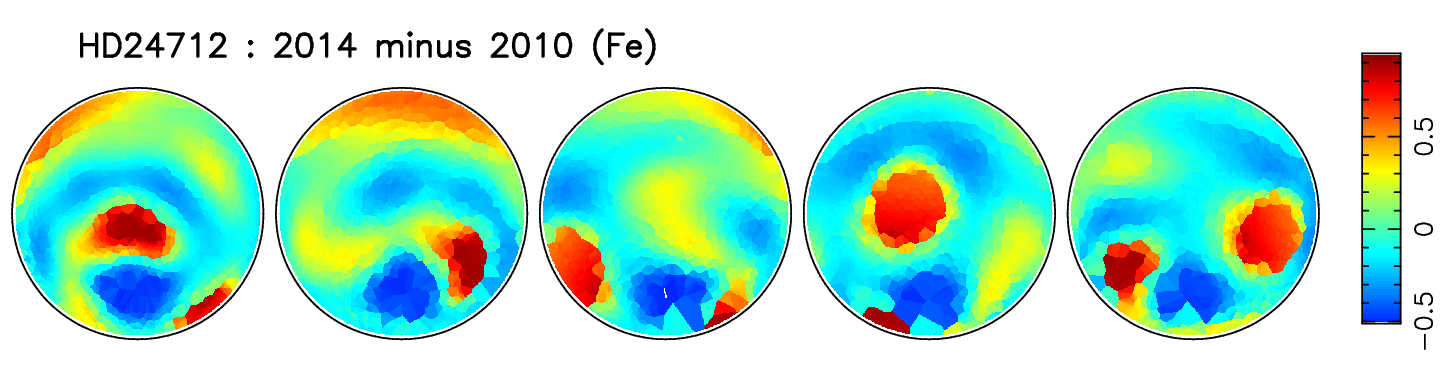}
\includegraphics[height=90mm, angle=270]{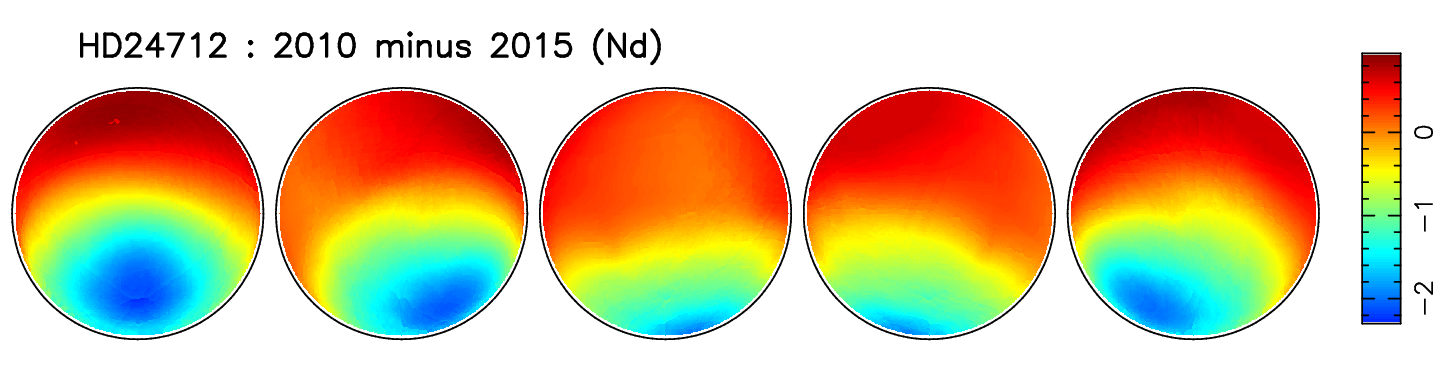}
\includegraphics[height=90mm, angle=270]{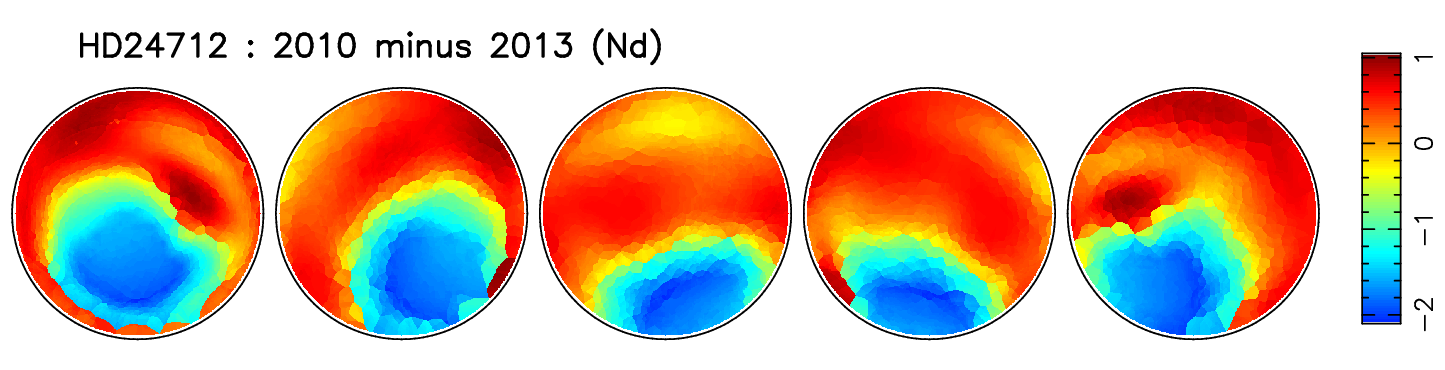}
\caption{
  ({\bf top 2}) HD\,24712 : Field strength differences between
  combinations of 3 maps at 5 equidistant phases.
  ({\bf middle 3}) Same as above, but for Fe abundances.
  ({\bf bottom 2}) Same as above, but for Cr abundances.
}
\label{HD242712_BFeNd_m}
\end{figure}

Using the same observations at 16 phases as in the 2014 paper, but
with $i = 120\degr$, \citet{RusomarovRuKoRyetal2015} arrived at yet
another magnetic topology and corresponding set of Fe and Nd abundance
maps. The magnetic field is no longer exclusively dipolar, but higher
order modes contribute only weakly (their Fig.\,5 and our
Fig.\ref{HD242712_B_m}). As to the new
Fe and Nd maps, the authors maintain that they \textit{``confirm''}
the 2010 findings and that they also show some \textit{``details''}
the previous study lacked. A look at Fig.\ref{HD242712_FeNd_m}
however reveals \textit{``details''} of more than 2\,dex in the Nd
maps, and of almost 1\,dex for Fe. No
mention is made of the huge differences with respect to the 2014 paper
or of this paper at all. No explanation is given for the large shift in
latitude of the Nd spot, the authors preferring a detailed discussion
of a possible marginal shift in longitude instead.

\begin{figure}
\includegraphics[height=90mm, angle=270]{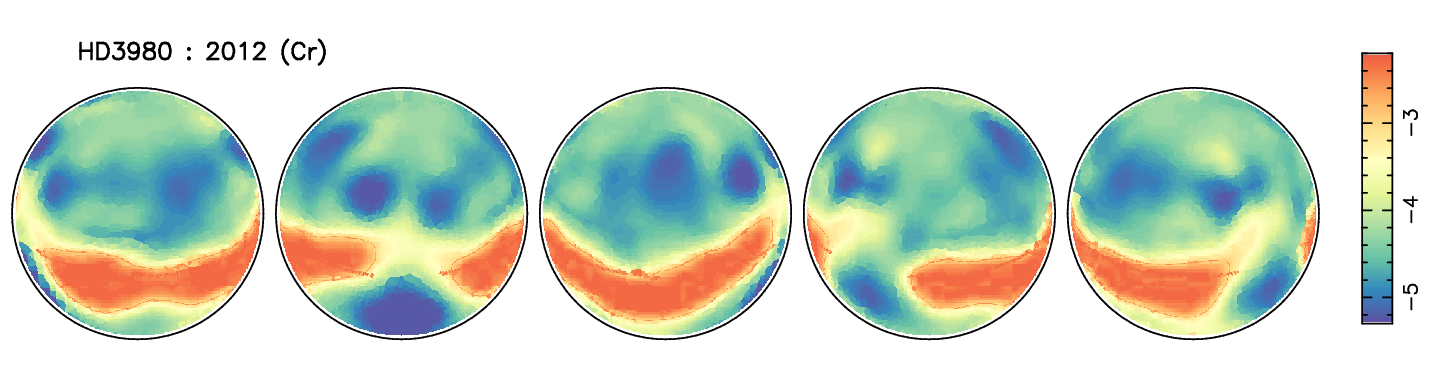}
\includegraphics[height=90mm, angle=270]{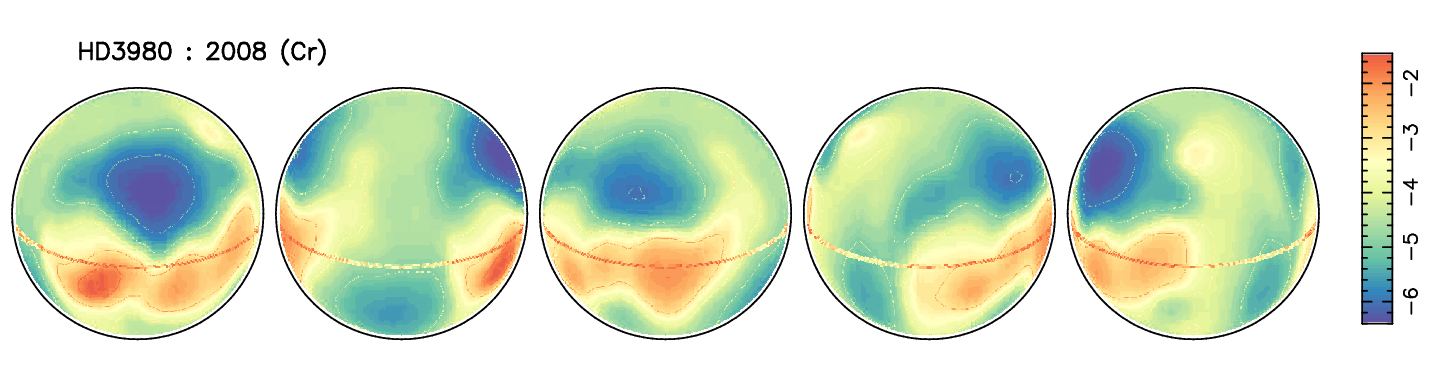}
\includegraphics[height=90mm, angle=270]{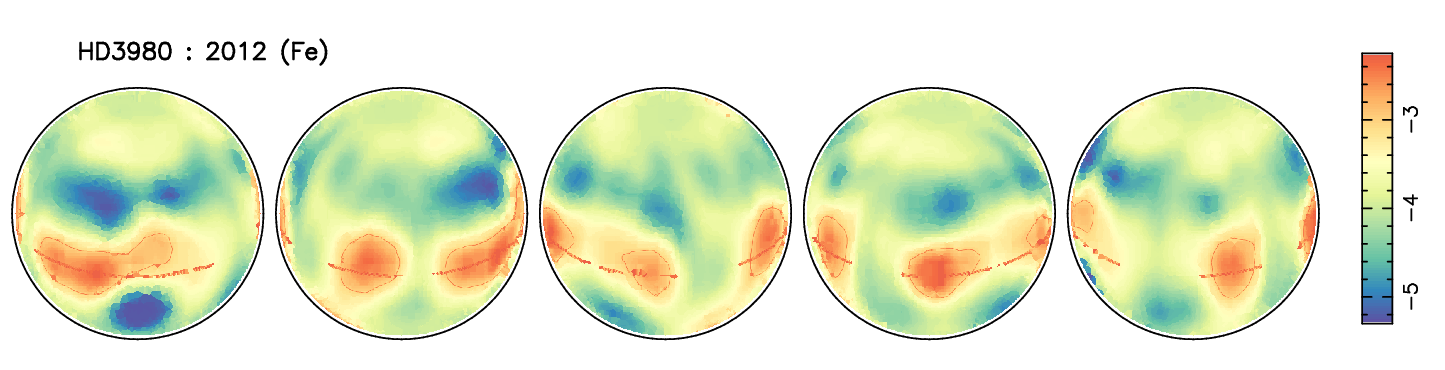}
\includegraphics[height=90mm, angle=270]{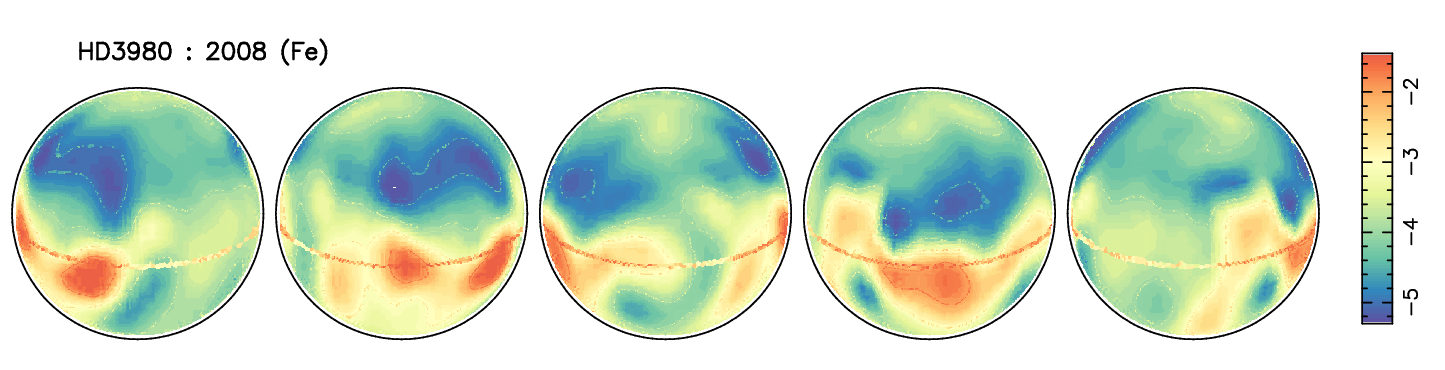}
\includegraphics[height=90mm, angle=270]{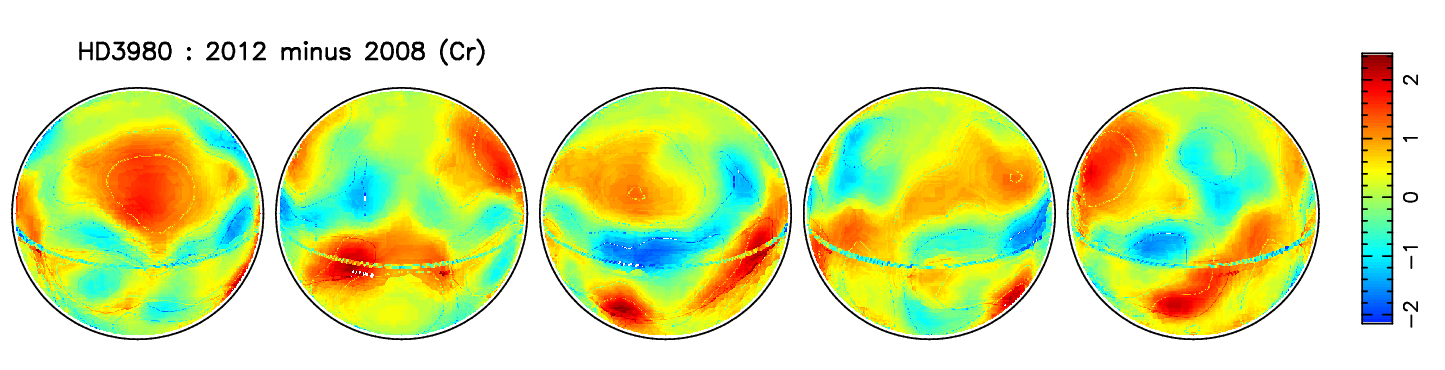}
\includegraphics[height=90mm, angle=270]{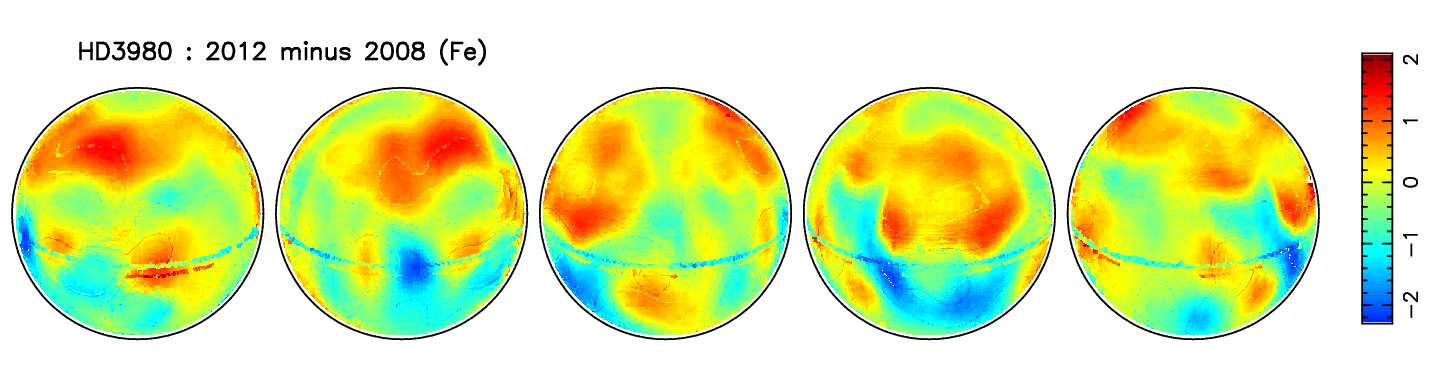}
\caption{
  ({\bf top 2})    HD\,3980 : maps of Cr, dated 2012 and 2008;
  ({\bf middle 2}) the same as above, but for Fe;
  ({\bf bottom 2}) abundance differences between the 2012 and 2008 maps.
}
\label{HD3980_FeNd_m}
\end{figure}

Given the low rotational velocity, there is very little difference in
the signal of a spot at $-30\degr$ and a spot much further to the south.
Looking at the quality (or the lack thereof) of the fits to the Stokes
profiles in 2014 as compared to 2015, one must not be surprised to
encounter multiple solutions to the inverse problem.

\begin{figure}
\includegraphics[height=90mm, angle=0]{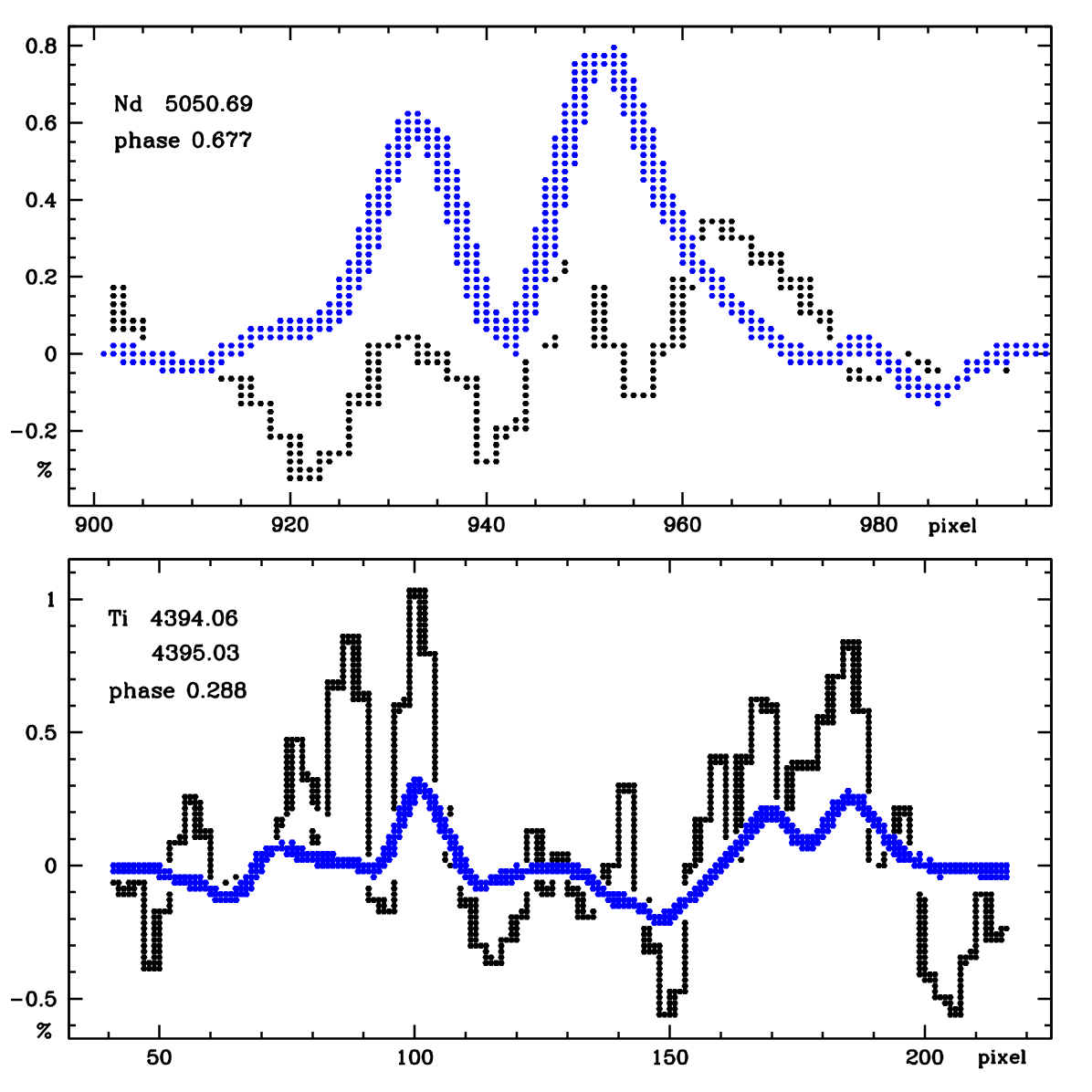}
\includegraphics[height=90mm, angle=0]{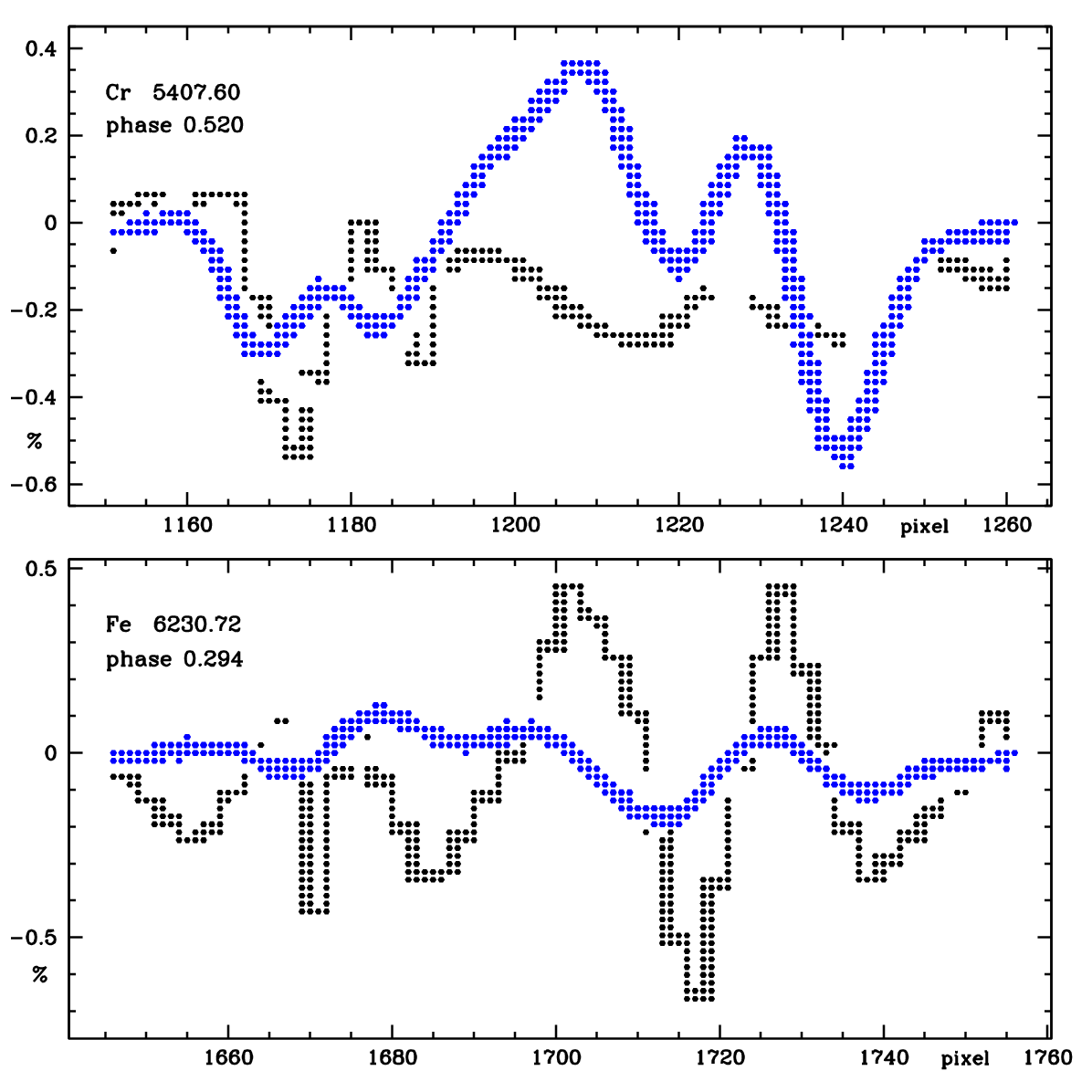}
\caption{
  49\,Cam : synthetic (blue) vs. observed (black) Stokes $Q$
  profiles of the elements Ti, Cr, Fe, and Nd according to
  \citet{SilvesterSiKoRuWa2017}.
}
\label{Q_49Cam}
\end{figure}

\subsection{HD\,3980}
\label{HD3980}

In a paper on Doppler mapping of HD\,3980, \citet{NesvacilNeLuShetal2012}
present abundance maps with stellar spots where silicon is as abundant
as hydrogen throughout the atmosphere, oxygen and manganese abundant as
helium. The intractable physical problems the existence of such spots would
imply have been discussed in some detail by \citet{StiftLeone2017b,
StiftLeone2017a}. Note that a magnetic field of an estimated 7\,kG polar
strength has not been taken into account in the DM inversion.

HD\,3980 would at first glance appear an unlikely object for our
error analysis, but there exists another set of Doppler maps presented
by \citet{Obbruggeretal2008} which allows us to carry out comparisons
between the respective Cr and Fe maps. The fits to the observed
profiles in the 2008 paper are of comparable quality to those in 2012,
but as Fig.\ref{HD3980_FeNd_m} reveals, the abundance maps do not agree
at all, differences reaching $\pm 2$\,dex. In the light of the manifest
physical inadequacies in the inversions and the conflicting Doppler maps,
there is no foundation to the claim made by \citet{NesvacilNeLuShetal2012}.
that ``a lack of up-to-date theoretical models'' is responsible for the
fact that no correlation is found between the abundance patterns in
HD\,3980 and predictions of diffusion theory.

\section{Conclusions}
\label{conclusions}

In view of the evidence gathered in the course of this investigation,
the gloomy outlook on ZDM as presented by \citet{StiftLeone2017a} is
corroborated:
\begin{itemize}
\item It is an indisputable fact that ZDM inversions admit of
  multiple solutions, even when built on exactly the same set
  of high-quality observations.
\item One has to accept that even magnetic topologies which
  are physically impossible -- in particular strong fields
  that do not obey the force-free condition -- can reproduce
  observed Stokes $IQUV$ profiles quite well. Unphysical
  solutions also suffer non-uniqueness.
\item The tests published by \citet{KochukhovKoPi2002} and
  \citet{Kochukhov2017} do not guarantee the plausibility
  of published ZDM maps, including the stars discussed above.
\item Not a single published test has explored inversions of truly
  complex magnetic geometries and horizontal abundance distributions.
  Tests never went beyond second-order magnetic topologies and
  idealised spots or rings.
\item In the presence of multiple maps that yield good fits to the
  observations, the choice between (widely) different horizontal
  abundance structures and magnetic geometries is quite arbitrary,
  often depending on the choice of lines and phases where substantial
  misfits are allowed. 49\,Cam is a telling example.
\end{itemize}
It follows that present-day ZDM results cannot
provide constraints of any kind to the modelling of atomic diffusion.
There remain also a few as yet unanswered questions:
\begin{itemize}
\item In ZDM solutions, can we really tolerate systematic discrepancies
  synthetic vs. observed profiles that attain levels by far exceeding
  the observational errors, as illustrated for 4 chemical elements in
  Fig.\ref{Q_49Cam}?
\item How does vertical stratification of chemical elements affect
  ZDM results? ``Putting diffusion theory to the test'' as done by
  \citet{KochRyab2018}, assumes the most perfectly symmetric magnetic
  geometry, a centred dipole with its axis in the equatorial plane of
  the star. What is the relevance of this test to stars such as
  $\alpha^2$\,CVn, 49\,Cam, 53\,Cam, HD\,32633, HD\,75049, HD\,119419,
  HD\,125428, HD\,133880?
\end{itemize}

\begin{acknowledgements}
  MJS expresses his gratitude to Dr.~S.~Bagnulo for having communicated second
  thoughts on ZDM, initiating a collaboration involving Armagh Observatory
  whose former director, Prof. M.~Bailey, generously provided unflinching
  support over many years. This work would never have been possible without
  the marvellous public GNAT Ada compiler of AdaCore.
\end{acknowledgements}

\bibliographystyle{aa}
\bibliography{ZDM}

\begin{thebibliography}{38}
\expandafter\ifx\csname natexlab\endcsname\relax\def\natexlab#1{#1}\fi

\bibitem[{{Braithwaite} \& {Spruit}(2017)}]{BraithSpruit2017}
{Braithwaite}, J. \& {Spruit}, H.~C. 2017, Royal Society Open Science, 4,
  160271

\bibitem[{{Donati} {et~al.}(2006){Donati}, {Howarth}, {Jardine}, {Petit},
  {Catala}, {Landstreet}, {Bouret}, {Alecian}, {Barnes}, {Forveille},
  {Paletou}, \& {Manset}}]{Donatietal2006}
{Donati}, J.~F., {Howarth}, I.~D., {Jardine}, M.~M., {et~al.} 2006, \mnras,
  370, 629

\bibitem[{{Jardine} {et~al.}(1999){Jardine}, {Barnes}, {Donati}, \& {Collier
  Cameron}}]{Jardineetal1999}
{Jardine}, M., {Barnes}, J.~R., {Donati}, J.-F., \& {Collier Cameron}, A. 1999,
  \mnras, 305, L35

\bibitem[{{Kochukhov}(2004)}]{Kochukhov2004IAUS224}
{Kochukhov}, O. 2004, in IAU Symposium, Vol. 224, The A-Star Puzzle, ed.
  J.~{Zverko}, J.~{Ziznovsky}, S.~J. {Adelman}, \& W.~W. {Weiss}, 433

\bibitem[{{Kochukhov}(2006)}]{Kochukhov2006}
{Kochukhov}, O. 2006, in Astronomical Society of the Pacific Conference Series,
  Vol. 358, Solar Polarization 4, ed. R.~{Casini} \& B.~W. {Lites}, 345

\bibitem[{{Kochukhov}(2017)}]{Kochukhov2017}
{Kochukhov}, O. 2017, \aap, 597, A58

\bibitem[{{Kochukhov} {et~al.}(2004{\natexlab{a}}){Kochukhov}, {Bagnulo},
  {Wade}, {Sangalli}, {Piskunov}, {Landstreet}, {Petit}, \&
  {Sigut}}]{KochukhovKoBaWaetal2004}
{Kochukhov}, O., {Bagnulo}, S., {Wade}, G.~A., {et~al.} 2004{\natexlab{a}},
  A\&A, 414, 613

\bibitem[{{Kochukhov} {et~al.}(2003){Kochukhov}, {Drake}, \& {de La
  Reza}}]{KochukhovKoDrRe2003}
{Kochukhov}, O., {Drake}, N.~A., \& {de La Reza}, R. 2003, in IAU Symposium,
  Vol. 210, Modelling of Stellar Atmospheres, ed. N.~{Piskunov}, W.~W. {Weiss},
  \& D.~F. {Gray}, D22

\bibitem[{{Kochukhov} {et~al.}(2004{\natexlab{b}}){Kochukhov}, {Drake},
  {Piskunov}, \& {de la Reza}}]{KochukhovKoDrPiRe2004}
{Kochukhov}, O., {Drake}, N.~A., {Piskunov}, N., \& {de la Reza}, R.
  2004{\natexlab{b}}, A\&A, 424, 935

\bibitem[{{Kochukhov} \& {Piskunov}(2002)}]{KochukhovKoPi2002}
{Kochukhov}, O. \& {Piskunov}, N. 2002, A\&A, 388, 868

\bibitem[{{Kochukhov} {et~al.}(2002){Kochukhov}, {Piskunov}, {Ilyin}, {Ilyina},
  \& {Tuominen}}]{KochukhovKoPiIlTu2002}
{Kochukhov}, O., {Piskunov}, N., {Ilyin}, I., {Ilyina}, S., \& {Tuominen}, I.
  2002, \aap, 389, 420

\bibitem[{{Kochukhov} \& {Ryabchikova}(2018)}]{KochRyab2018}
{Kochukhov}, O. \& {Ryabchikova}, T.~A. 2018, \mnras, 474, 2787

\bibitem[{{Kochukhov} {et~al.}(2017){Kochukhov}, {Silvester}, {Bailey},
  {Landstreet}, \& {Wade}}]{KochukhovKoSiBaLaWa2017}
{Kochukhov}, O., {Silvester}, J., {Bailey}, J.~D., {Landstreet}, J.~D., \&
  {Wade}, G.~A. 2017, \aap, 605, A13

\bibitem[{{Kochukhov} \& {Wade}(2010)}]{KochukhovKoWa2010}
{Kochukhov}, O. \& {Wade}, G.~A. 2010, A\&A, 513, A13

\bibitem[{{L{\"u}ftinger} {et~al.}(2010){L{\"u}ftinger}, {Kochukhov},
  {Ryabchikova}, {Piskunov}, {Weiss}, \& {Ilyin}}]{LuftingerLuKoRyetal2010b}
{L{\"u}ftinger}, T., {Kochukhov}, O., {Ryabchikova}, T., {et~al.} 2010, A\&A,
  509, A71

\bibitem[{{Nesvacil} {et~al.}(2012){Nesvacil}, {L{\"u}ftinger}, {Shulyak},
  {Obbrugger}, {Weiss}, {Drake}, {Hubrig}, {Ryabchikova}, {Kochukhov},
  {Piskunov}, \& {Polosukhina}}]{NesvacilNeLuShetal2012}
{Nesvacil}, N., {L{\"u}ftinger}, T., {Shulyak}, D., {et~al.} 2012, A\&A, 537,
  A151

\bibitem[{{Obbrugger} {et~al.}(2008){Obbrugger}, {L{\"u}ftinger}, {Nesvacil},
  {Kochukhov}, \& {Weiss}}]{Obbruggeretal2008}
{Obbrugger}, M., {L{\"u}ftinger}, T., {Nesvacil}, N., {Kochukhov}, O., \&
  {Weiss}, W.~W. 2008, Contributions of the Astronomical Observatory Skalnate
  Pleso, 38, 347

\bibitem[{{Piskunov}(2008)}]{Piskunov2008}
{Piskunov}, N. 2008, Physica Scripta Volume T, 133, 014017

\bibitem[{{Piskunov} \& {Wehlau}(1990)}]{PiskunovWehlau1990}
{Piskunov}, N.~E. \& {Wehlau}, W.~H. 1990, \aap, 233, 497

\bibitem[{{Rusomarov} {et~al.}(2018){Rusomarov}, {Kochukhov}, \&
  {Lundin}}]{RusomarovRuKoLu2018}
{Rusomarov}, N., {Kochukhov}, O., \& {Lundin}, A. 2018, \aap, 609, A88

\bibitem[{{Rusomarov} {et~al.}(2014){Rusomarov}, {Kochukhov}, \&
  {Piskunov}}]{Rusomarovetal2014}
{Rusomarov}, N., {Kochukhov}, O., \& {Piskunov}, N. 2014, in IAU Symposium,
  Vol. 302, Magnetic Fields throughout Stellar Evolution, ed. P.~{Petit},
  M.~{Jardine}, \& H.~C. {Spruit}, 304--305

\bibitem[{{Rusomarov} {et~al.}(2016){Rusomarov}, {Kochukhov}, {Ryabchikova}, \&
  {Ilyin}}]{RusomarovRuKoRyIl2016}
{Rusomarov}, N., {Kochukhov}, O., {Ryabchikova}, T., \& {Ilyin}, I. 2016, \aap,
  588, A138

\bibitem[{{Rusomarov} {et~al.}(2015){Rusomarov}, {Kochukhov}, {Ryabchikova}, \&
  {Piskunov}}]{RusomarovRuKoRyetal2015}
{Rusomarov}, N., {Kochukhov}, O., {Ryabchikova}, T., \& {Piskunov}, N. 2015,
  A\&A, 573, A123

\bibitem[{{Semel}(1989)}]{SemelSe1989}
{Semel}, M. 1989, A\&A, 225, 456

\bibitem[{{Silvester} {et~al.}(2017){Silvester}, {Kochukhov}, {Rusomarov}, \&
  {Wade}}]{SilvesterSiKoRuWa2017}
{Silvester}, J., {Kochukhov}, O., {Rusomarov}, N., \& {Wade}, G.~A. 2017,
  \mnras, 471, 962

\bibitem[{{Silvester} {et~al.}(2014{\natexlab{a}}){Silvester}, {Kochukhov}, \&
  {Wade}}]{SilvesterSiKoWa2014a}
{Silvester}, J., {Kochukhov}, O., \& {Wade}, G.~A. 2014{\natexlab{a}}, \mnras,
  440, 182

\bibitem[{{Silvester} {et~al.}(2014{\natexlab{b}}){Silvester}, {Kochukhov}, \&
  {Wade}}]{SilvesterSiKoWa2014b}
{Silvester}, J., {Kochukhov}, O., \& {Wade}, G.~A. 2014{\natexlab{b}}, MNRAS,
  444, 1442

\bibitem[{{Silvester} {et~al.}(2015){Silvester}, {Kochukhov}, \&
  {Wade}}]{SilvesterSiKoWa2015}
{Silvester}, J., {Kochukhov}, O., \& {Wade}, G.~A. 2015, \mnras, 453, 2163

\bibitem[{{Silvester} {et~al.}(2011){Silvester}, {Wade}, {Kochukhov},
  {Landstreet}, \& {Bagnulo}}]{SilvesterWaKoLaBa2011}
{Silvester}, J., {Wade}, G., {Kochukhov}, O., {Landstreet}, J., \& {Bagnulo},
  S. 2011, in Astronomical Society of the Pacific Conference Series, Vol. 449,
  Astronomical Polarimetry 2008: Science from Small to Large Telescopes, ed.
  P.~{Bastien}, N.~{Manset}, D.~P. {Clemens}, \& N.~{St-Louis}, 280

\bibitem[{Silvester {et~al.}(2007)Silvester, Wade, Kochukhov, Landstreet, \&
  Bagnulo}]{Silvester2007magneticdopplerimagingap}
Silvester, J., Wade, G.~A., Kochukhov, O., Landstreet, J.~D., \& Bagnulo, S.
  2007, Magnetic Doppler Imaging of Ap stars

\bibitem[{{Silvester} {et~al.}(2008){Silvester}, {Wade}, {Kochukhov},
  {Landstreet}, \& {Bagnulo}}]{SilvesterWaKoLaBa2008}
{Silvester}, J., {Wade}, G.~A., {Kochukhov}, O., {Landstreet}, J.~D., \&
  {Bagnulo}, S. 2008, Contributions of the Astronomical Observatory Skalnate
  Pleso, 38, 341

\bibitem[{{Stift} \& {Leone}(2017{\natexlab{a}})}]{StiftLeone2017b}
{Stift}, M.~J. \& {Leone}, F. 2017{\natexlab{a}}, \mnras, 465, 2880

\bibitem[{{Stift} \& {Leone}(2017{\natexlab{b}})}]{StiftLeone2017a}
{Stift}, M.~J. \& {Leone}, F. 2017{\natexlab{b}}, \apj, 834, 24

\bibitem[{{Stift} \& {Leone}(2022)}]{StiftLeone2022}
{Stift}, M.~J. \& {Leone}, F. 2022, \aap, 659, A33

\bibitem[{{Stift} {et~al.}(2012){Stift}, {Leone}, \&
  {Cowley}}]{StiftStLeCo2012}
{Stift}, M.~J., {Leone}, F., \& {Cowley}, C.~R. 2012, MNRAS, 419, 2912

\bibitem[{{Tikhonov} \& {Goncharsky}(1987)}]{Tikhonov1987}
{Tikhonov}, A.~N. \& {Goncharsky}, A.~V. 1987, {Ill-posed problems in the
  natural sciences}

\bibitem[{{Vogt} {et~al.}(1987){Vogt}, {Penrod}, \& {Hatzes}}]{VogtVoPeHa1987}
{Vogt}, S.~S., {Penrod}, G.~D., \& {Hatzes}, A.~P. 1987, ApJ, 321, 496

\bibitem[{{Winch} {et~al.}(2005){Winch}, {Ivers}, {Turner}, \&
  {Stening}}]{WinchWiIvTuSt2005}
{Winch}, D.~E., {Ivers}, D.~J., {Turner}, J.~P.~R., \& {Stening}, R.~J. 2005,
  Geophysical Journal International, 160, 487

\end{thebibliography}

\end{document}